\documentclass[12pt]{article}

\usepackage{amssymb, amsmath, amsthm, amsfonts}
\usepackage{multirow}
\usepackage{booktabs}
\usepackage{subfigure}
\usepackage[authoryear]{natbib}
\usepackage[colorlinks,citecolor=blue,urlcolor=blue]{hyperref}
\usepackage{graphicx}
\usepackage{newtxtext}
\usepackage{epsfig,latexsym,verbatim}
\usepackage{bm}
\usepackage{mathrsfs}
\usepackage{multicol}
\usepackage{enumitem}
\usepackage{makecell}
\usepackage[figuresright]{rotating}
\usepackage{xr}
\graphicspath{{Fig/}}

\usepackage{booktabs}
\usepackage{subfigure}
\usepackage{graphics}
\usepackage{lmodern}
\usepackage{bm}
\usepackage{tikz-cd}
\usepackage{comment}
\usepackage{algorithm,algpseudocode}
\usepackage{algorithmicx}
\usepackage{appendix}
\usepackage{float}
\newfloat{algorithm}{t}{lop}
\newcommand{\blind}{0}

\theoremstyle{plain}

\newtheorem{theorem}{Theorem}

\newtheorem{assumption}{Assumption}

\newtheorem{remark}{Remark}
\newtheorem{corollary}{Corollary}
\theoremstyle{remark}
\newtheorem{definition}{Definition}

\newcommand{\DD}{{\mathcal D}}

\newcommand{\m}{{\mathcal M}}

\newcommand{\N}{{\mathcal N}}

\newcommand{\PP}{{\mathbb P}}

\newcommand{\R}{{\mathbb R}}

\newcommand{\Exp}{{\operatorname{Exp}}}
\newcommand{\Log}{{\operatorname{Log}}}
\newcommand{\AI}{{\operatorname{AI}}}

\newcommand{\ddp}{{\operatorname{dp}}}
\newcommand{\Hess}{{\operatorname{Hess}}}
\newcommand{\cov}{{\operatorname{Cov}}}

\newcommand{\vecd}{{\operatorname{vecd}}}

\newcommand{\RG}{\operatorname{RG}}
\newcommand{\EWG}{\operatorname{EWG}}

\newcommand{\Oop}{{\mathcal{O}_{\mathbb{P}}}}

\begin{document}

\def\spacingset#1{\renewcommand{\baselinestretch}%
{#1}\small\normalsize} \spacingset{1}

\date{}
\if0\blind
{
  \title{\bf Differentially private inference framework for Riemannian manifold data}
  \author{Yangdi Jiang \footnotemark[1] $\,$\footnotemark[3],$\,$ Xiaotian Chang\footnotemark[1]$\,$ \footnotemark[3], $\,$ and$\,$ Qirui Hu\footnotemark[2]$\,$ \footnotemark[4] \hspace{.2cm}}
  \maketitle
  \renewcommand{\thefootnote}{\fnsymbol{footnote}}

  \footnotetext[1]{School of Physical and Mathematical Sciences, Nanyang Technological University}
  \footnotetext[2]{School of Statistics and Data Science, Shanghai University of Finance and Economics}
  \footnotetext[3]{Co-first authors. These authors contributed equally to this work.}
  \footnotetext[4]{Corresponding author: huqirui@mail.shufe.edu.cn}
}

\if1\blind
{
  \bigskip
  \bigskip
  \bigskip
  \begin{center}
    { \bf Differentially private inference framework for Riemannian manifold data}
  \end{center}
  \medskip
} \fi

\bigskip

\abstract{We propose a novel and systematic differentially private (DP) inference framework for non-Euclidean data. First, we design two types of DP mechanisms for the Fr\'echet mean and variance for i.i.d. Riemannian manifold-valued data, tailored to different geometric structures and accompanied by analytic privacy budgets calibrated to the geometry of the underlying manifold. Second, we establish the consistency and central limit theorems (CLTs) of the proposed DP estimators, enabling a suite of statistical inference procedures under privacy constraints. Furthermore, we provide comprehensive implementation guidelines and feasible procedures, including consistent DP estimators of the asymptotic variance in the CLTs. Extensive numerical experiments support the proposed methodologies. Finally, we demonstrate the effectiveness of our approach on real-world medical image and sociological datasets supported on two representative manifolds.}

\vspace{0.5cm}

\newpage
\spacingset{1.9}

\section{Introduction}
\begin{figure}[h]
    \centering
    \subfigure[Earnings]{\includegraphics[width=0.24\textwidth, trim=3cm 3cm 3cm 3cm, clip]{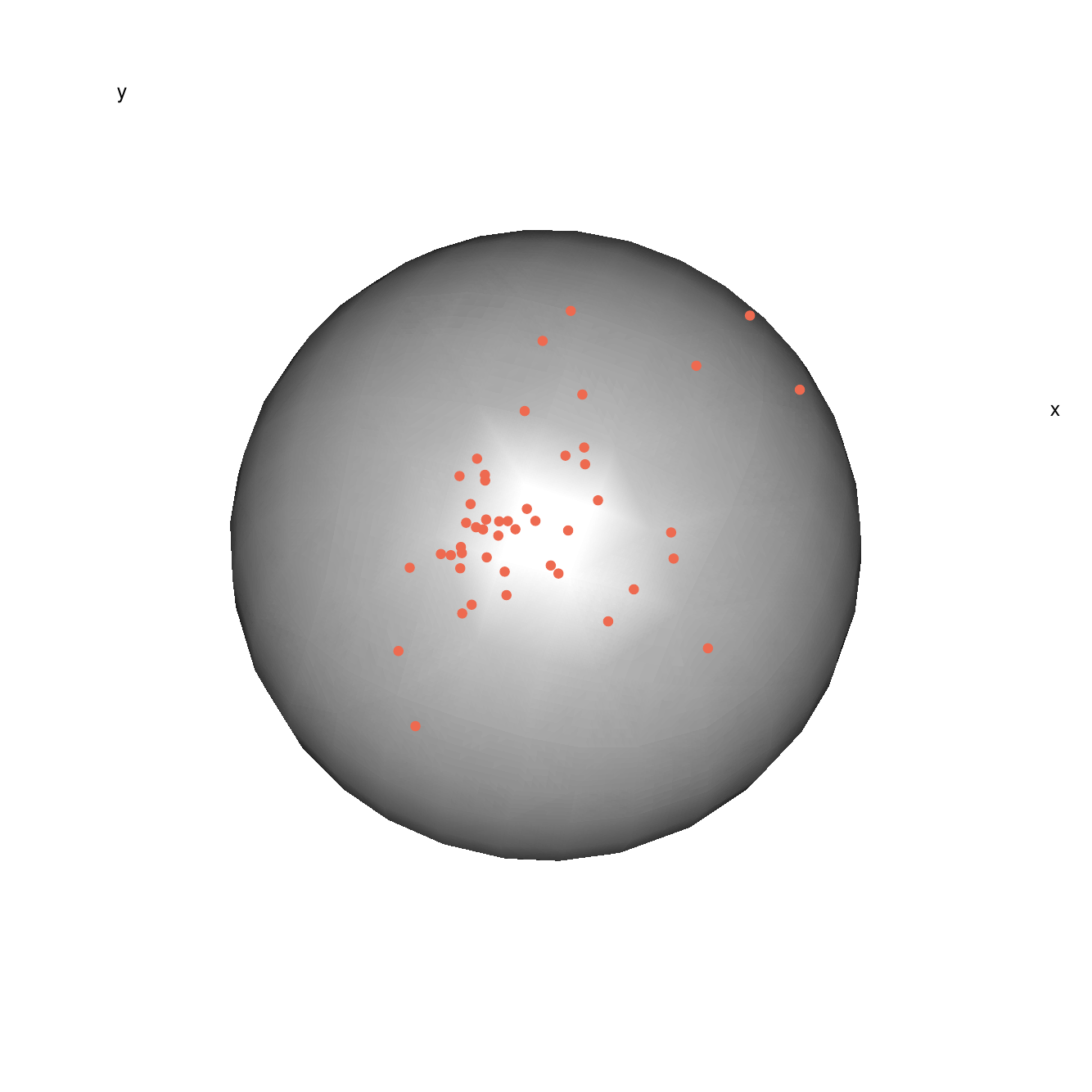}}
    \subfigure[Social status]{\includegraphics[width=0.24\textwidth, trim=3cm 3cm 3cm 3cm, clip]{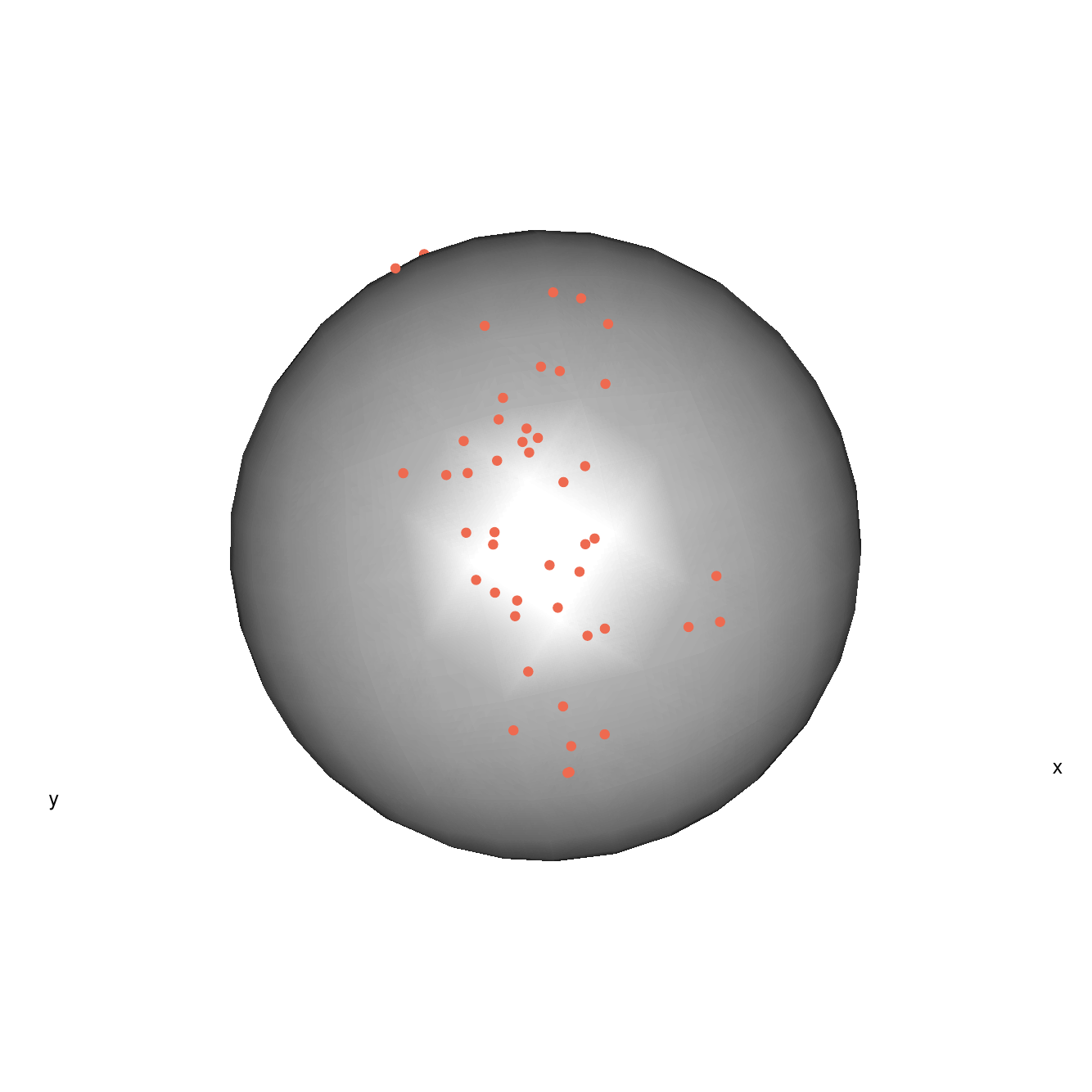}}
    \subfigure[Reward]{\includegraphics[width=0.24\textwidth, trim=3cm 3cm 3cm 3cm, clip]{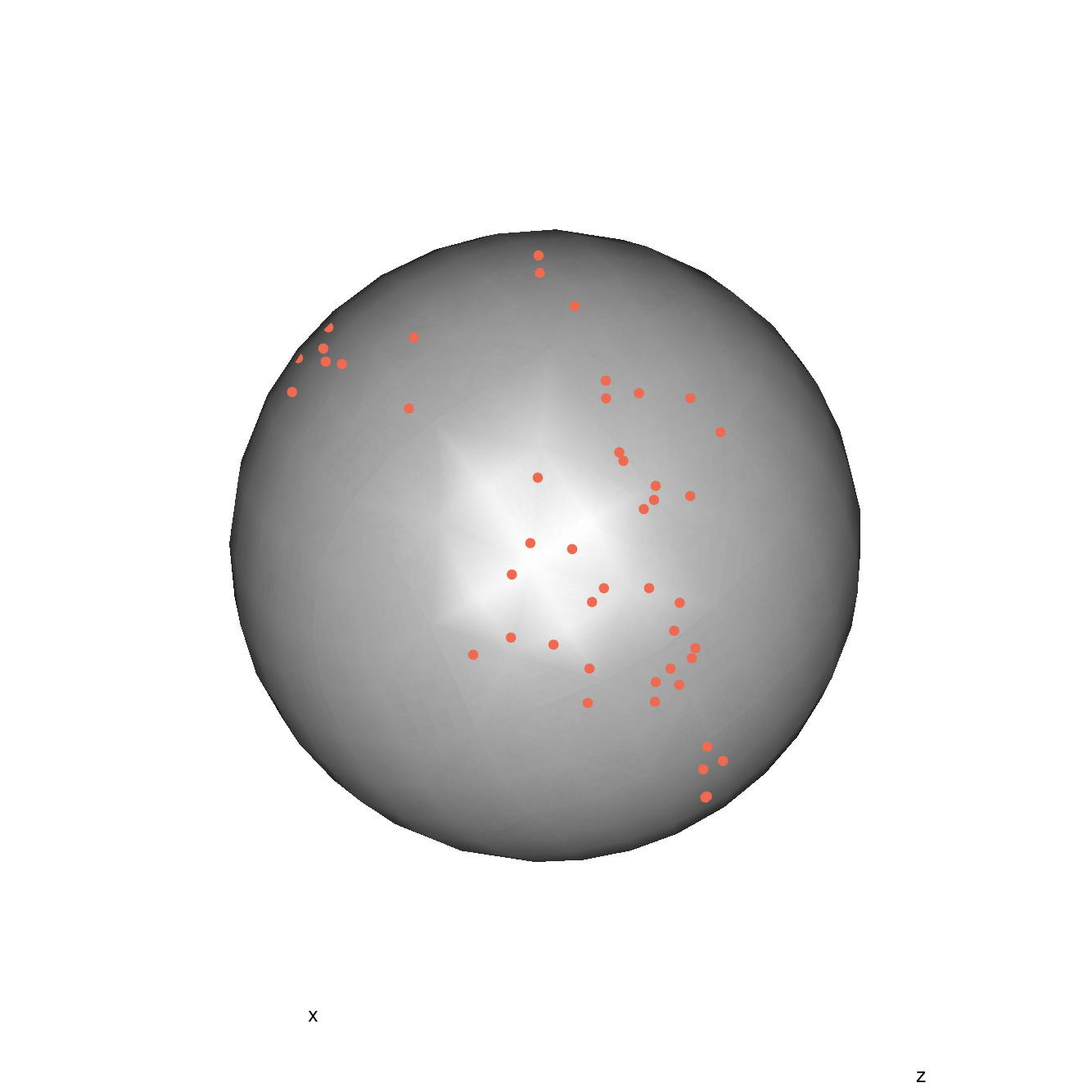}}
    \subfigure[Social usefulness]{\includegraphics[width=0.24\textwidth, trim=3cm 3cm 3cm 3cm, clip]{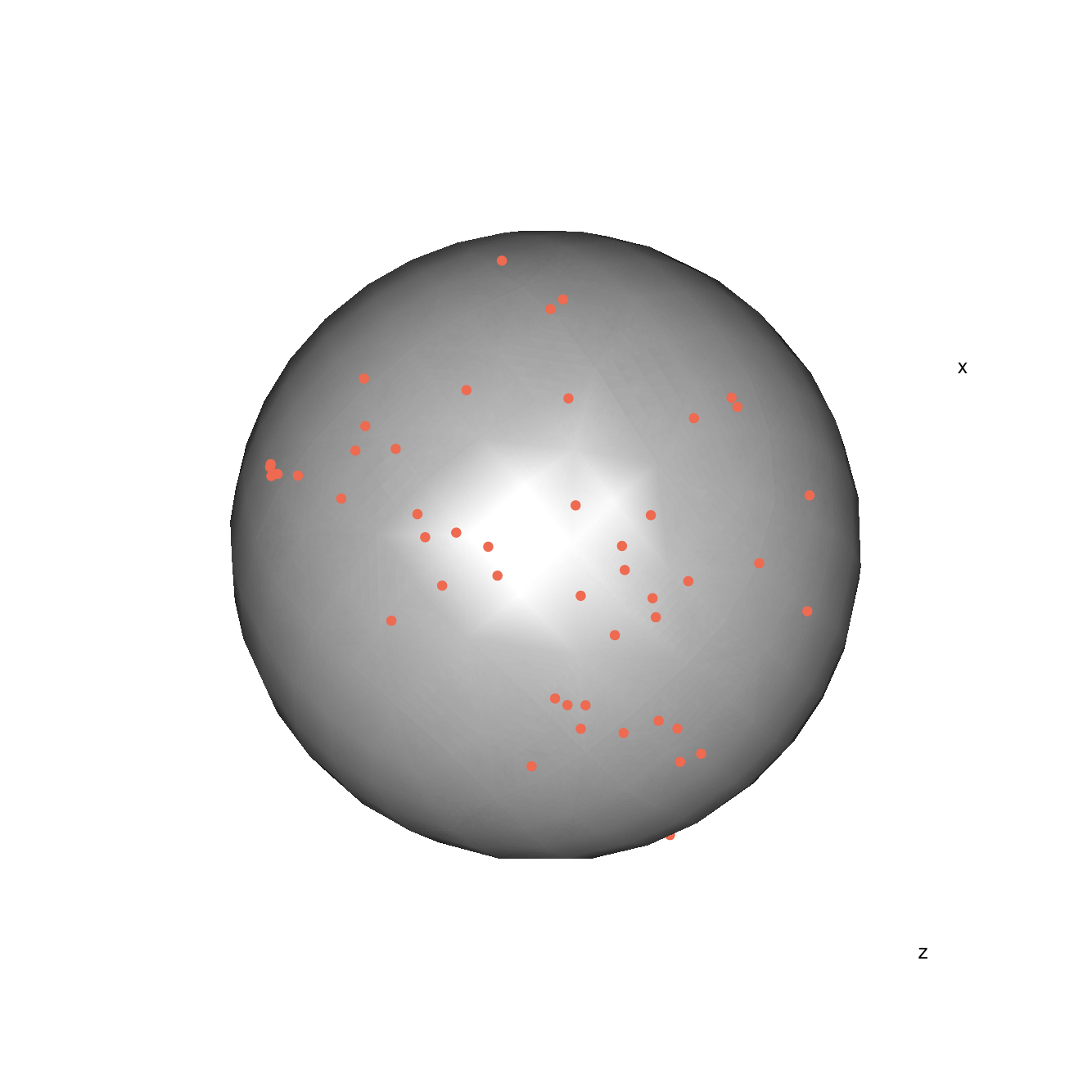}}
    \caption{Occupational judgment data based on four criteria
    \citep[p.~301]{fisher1987spherical}.}
    \label{fig:occupation}
\end{figure}

\begin{figure}[h]
    \centering
    \subfigure[Class 0]{\includegraphics[width=0.24\textwidth]{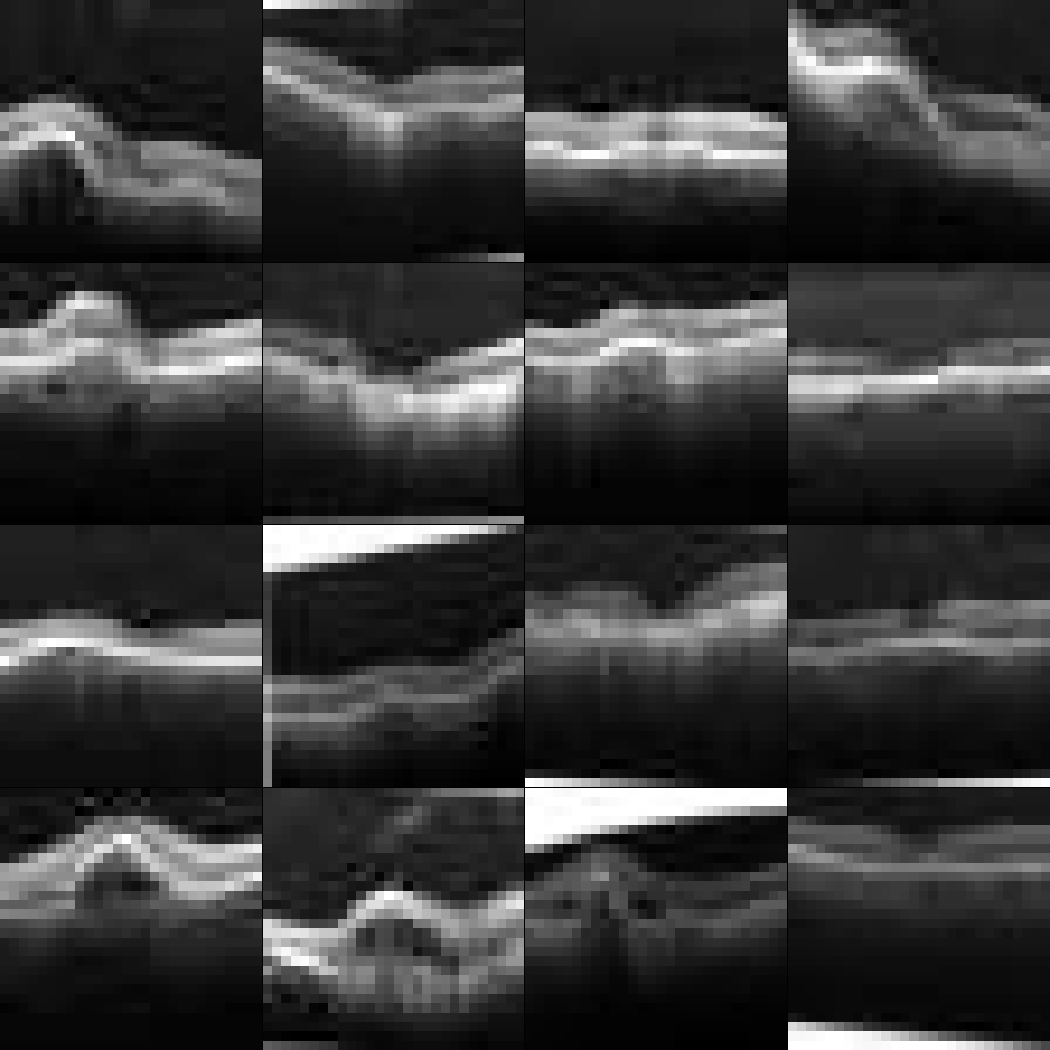}}
    \subfigure[Class 1]{\includegraphics[width=0.24\textwidth]{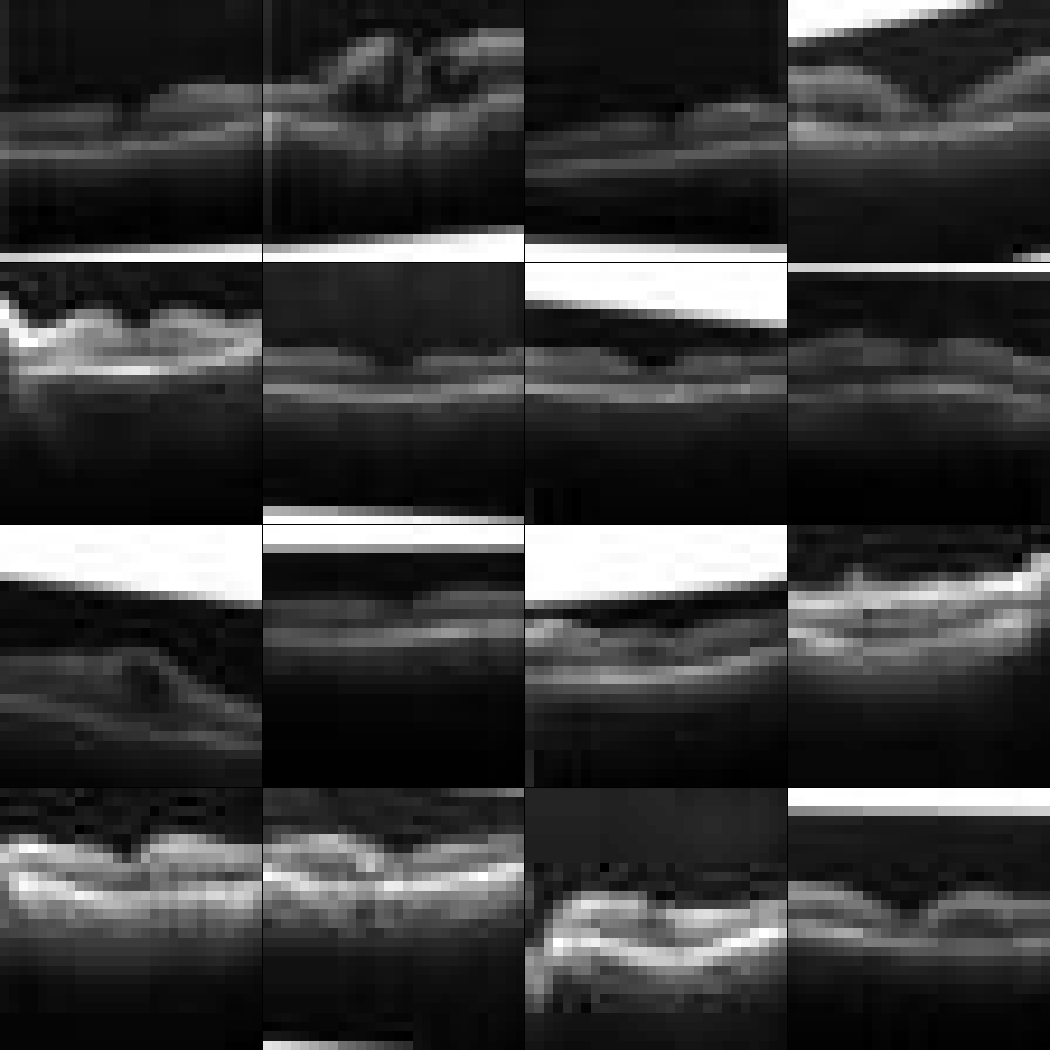}}
    \subfigure[Class 2]{\includegraphics[width=0.24\textwidth]{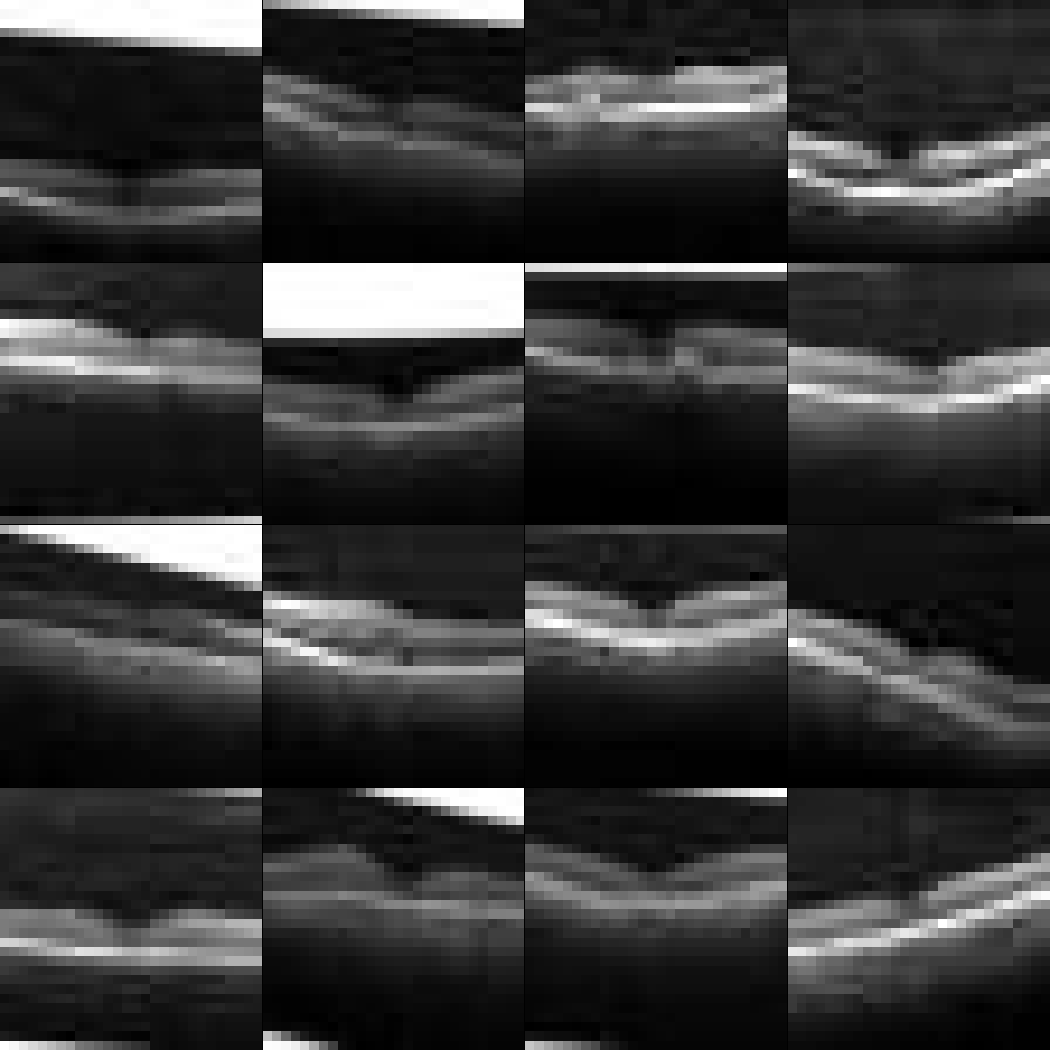}}
    \subfigure[Class 3]{\includegraphics[width=0.24\textwidth]{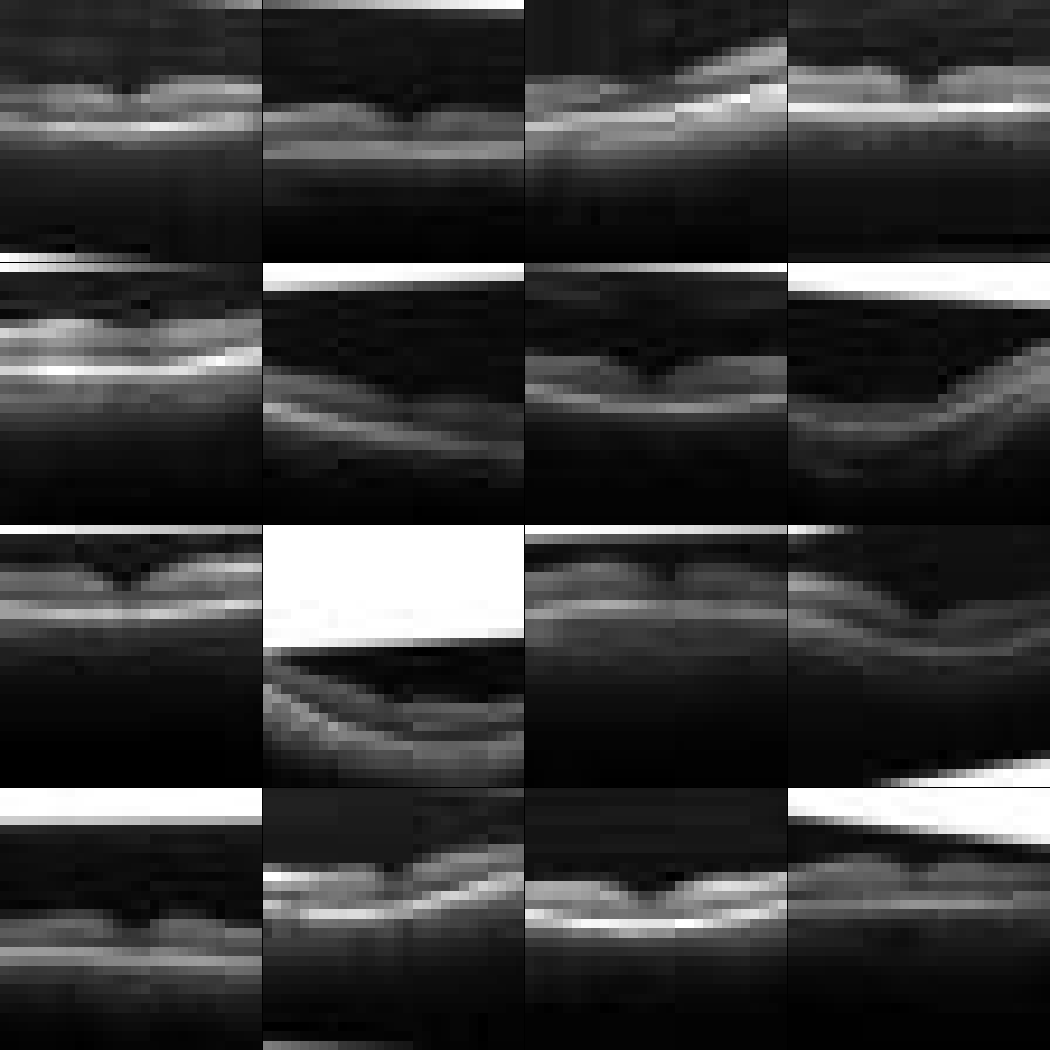}}
    \caption{Representative images from four diagnosis categories in the
    \textsf{OCTMNIST} data set
    \citep{Yang2021MedMNISTV1,Yang2021MedMNISTV2}.}
    \label{fig:octmnist}
\end{figure}

Many contemporary statistical problems involve observations whose natural
sample space is not a vector space. Symmetric positive definite matrices
(SPDMs), for example, arise as covariance descriptors in imaging analysis
\citep{tuzel2006covariance}; in diffusion tensor imaging, they represent the
covariance structure of water-molecule diffusion
\citep{basser1994mr,pennec2019riemannian}. Directional observations lie on
spheres and form the basis of directional statistics
\citep{fisher1987spherical,mardia2000directional}. For such data, ordinary
Euclidean operations may violate structural constraints or distort the
scientifically relevant notion of distance. Fr\'{e}chet analysis provides an
intrinsic alternative by defining location and variation through the geodesic
distance. Figures~\ref{fig:occupation} and~\ref{fig:octmnist} display two
motivating examples used later in the paper: occupational judgment data on a
sphere and covariance descriptors extracted from medical images.

The same applications that motivate intrinsic analysis also raise privacy
concerns. Medical images, covariance descriptors, directional records and
social-science measurements may encode individual-level information that
should not be disclosed through published statistics or shared data sets.
Differential privacy (DP) offers a formal way to release statistical
summaries while controlling the information revealed about any single
individual \citep{dwork2006a,dwork2006b}. For Euclidean vector-valued
queries, a common strategy is to add random noise calibrated to sensitivity,
an idea underlying Gaussian and related mechanisms and much of the recent
literature on DP statistical inference
\citep{Wasserman2010,Balle2018ImprovingTG,dong2022gaussian,liu2023online,liutuning}.
This strategy does not transfer directly to Riemannian manifolds. A private
Fr\'{e}chet mean must remain on the manifold, its sensitivity is measured by
geodesic distance and affected by curvature, and valid inference must account
simultaneously for sampling variability, privacy noise and the local geometry
of exponential and logarithm maps.

Recent work has begun to address privacy for non-Euclidean data, including
sensitivity analysis and private mechanisms on Riemannian manifolds
\citep{reimherr2021,jiang2023gaussian}, as well as private Fr\'{e}chet mean
estimation for SPDMs under the log-Euclidean metric \citep{Saiteja2023}.
These results provide important building blocks, but a general inference
framework for privatized Fr\'{e}chet means and variances remains unavailable
for broad classes of positively curved and nonpositively curved manifolds.
Existing Riemannian Gaussian calibration can also be implicit and
computationally expensive, while the effect of privacy noise on confidence
regions and intrinsic variance estimation has not been systematically
developed.

We study i.i.d. confidential observations
$D=\{X_1,\ldots,X_n\}$ taking values on a $d$-dimensional Riemannian manifold
$\m$. The parameters of interest are the population Fr\'{e}chet mean $\eta$
and the Fr\'{e}chet variance $V$, defined in
Definition~\ref{DEF:ferchetmean}. The privacy framework is Gaussian
differential privacy (GDP), which uses a single privacy parameter $\mu$ and is
analytically connected to $(\epsilon,\delta)$-DP \citep{dong2022gaussian}.
The proposed procedures release differentially private versions of the
Fr\'{e}chet mean and variance and use their limiting distributions to form
confidence regions and confidence intervals.

The construction is designed to respect the geometry of the sample space. For
homogeneous manifolds with positive curvature, we use a Riemannian Gaussian
mechanism. For Hadamard manifolds, we use an exponential-wrapped Gaussian
mechanism, which perturbs a tangent-space representation and maps the result
back to the manifold through the exponential map. In both cases, the released
mean remains manifold-valued, and the privacy scale is calibrated through the
geodesic sensitivity of the sample Fr\'{e}chet mean. For the positive-curvature
homogeneous case, this calibration yields an analytic relationship among
sensitivity, curvature and the GDP parameter. It generalizes the classical
Euclidean calibration principle, relaxes constant-curvature requirements and
avoids numerical inversion of the privacy budget required by some existing
Riemannian Gaussian implementations. The Fr\'{e}chet variance is privatized
through a calibrated Euclidean Gaussian release, so that both intrinsic
location and intrinsic dispersion can be released with privacy protection.

The inferential theory requires more than adding an independent Gaussian term
to a Euclidean estimator. On a manifold there is no global vector addition,
the privacy perturbation is generated through geometric maps, and the
normalized released estimator is represented only in local coordinates. The
analysis therefore combines the intrinsic $M$-estimation expansion of the
Fr\'{e}chet functional with a local expansion of the privacy mechanism. This
requires control of exponential and logarithm maps, curvature-induced
remainder terms and the way geodesic sensitivity enters the limiting
covariance. Under the stated regularity conditions, the resulting central
limit theorems separate intrinsic sampling variation from the additional
privacy-noise contribution. This decomposition leads to differentially private
confidence regions for the Fr\'{e}chet mean and differentially private
confidence intervals for the Fr\'{e}chet variance.

The framework also accounts for nonstandard geometric asymptotics. In
manifold-valued inference, the Fr\'{e}chet functional may be flat or nearly
flat around its minimizer, producing smeariness and convergence rates slower
than the usual $n^{-1/2}$ rate. We derive the corresponding smeary
asymptotics for the privatized Fr\'{e}chet mean, clarifying how privacy noise
interacts with geometry-induced slow convergence. This identifies regimes in
which the geometry of the parameter space, rather than the privacy mechanism
alone, determines the limiting behavior.

The methodology is accompanied by implementable covariance estimation and
finite-sample evaluation. We construct consistent differentially private
estimators for the covariance components appearing in the limiting
distributions, provide practical algorithms for the proposed private releases,
and examine coverage under different privacy budgets. Simulations and
real-data analyses illustrate the methods on spherical occupational judgment
data and SPDM summaries derived from \textsf{OCTMNIST}. These results turn
privacy-preserving manifold-valued estimation into a full inference procedure,
with uncertainty quantification for both intrinsic location and intrinsic
dispersion.

The paper connects intrinsic inference for manifold-valued data with
differentially private statistical methodology. Foundational large-sample
theory for intrinsic and extrinsic Fr\'{e}chet means was developed by
\citet{Bhattacharya2003,Bhattacharya2005}; later work addressed smeariness
\citep{smearyCLT2019}, Fr\'{e}chet analysis of variance and regression
\citep{dubey2019frechet,petersen2019frechet}, change-point analysis
\citep{dubey2020frechet}, functional random objects
\citep{dubey2020functional,lin2021functional,shao2022intrinsic}, and robust,
generalized and high-dimensional extensions
\citep{yang2020shrinkage,mccormack2023equivariant,chakraborty2019statistics,
evans2024limit,dubey2024metric,lee2024huber,blanchard2025frechet}. Most
inferential theory for DP, by contrast, remains Euclidean. The present work
bridges these areas by combining formal privacy guarantees with intrinsic
Fr\'{e}chet inference while preserving the geometry of the released
estimators.

The paper is organized as follows. Section~\ref{sec2} reviews Riemannian
preliminaries, Fr\'{e}chet summaries and differential privacy.
Section~\ref{sec3} presents the private mechanisms and asymptotic theory.
Section~\ref{sec4} gives implementation details, including estimation of the
limiting variance components under privacy constraints. Sections~\ref{sec5}
and~\ref{sec6} report simulations and real-data applications. Additional
geometric background, regularity conditions, smeary asymptotics and technical
proofs are provided in the supplementary material.

\section{Preliminaries}\label{sec2}

\subsection{Basic Settings}
To facilitate our presentation throughout this article, we introduce the following notations and definitions. Let $\mathcal{M}$ denote a $d$-dimensional Riemannian manifold equipped with a Riemannian metric $g$, which induces a geodesic distance function $\rho: \mathcal{M} \times \mathcal{M} \to \mathbb{R}$.\footnote{We assume that $g$ is $C^\infty$-smooth.} We denote by $\exp_p : T_p\mathcal{M} \to \mathcal{M}$ the exponential map at a point $p \in \mathcal{M}$, which sends a tangent vector $\mathbf{v} \in T_p\mathcal{M}$ to the point reached at unit time by the geodesic starting from $p$ with initial velocity $\mathbf{v}$. While $\exp_p$ is a local diffeomorphism near the origin, it is not globally injective; hence, multiple geodesics may connect two points $p, q \in \mathcal{M}$. The cut locus of $p$, denoted by $C(p)$, is the set of points where $\exp_p$ ceases to be minimizing or injective. Equivalently, $C(p)$ consists of points connected to $p$ by more than one minimizing geodesic. On $\mathcal{M} \setminus C(p)$, the exponential map remains a smooth diffeomorphism, and its inverse is well defined. We denote this inverse by $\log_p$, following standard convention. Furthermore, we assume that $\mathcal{M}$ is geodesically complete, which ensures that $(\mathcal{M}, \rho)$ forms a complete metric space. The remaining notations are summarized in Table~\ref{tab:notation} in Appendix~A.

\subsection{Fr\'echet Mean on Riemannian Manifold}\label{sec_manifold}

\begin{definition}\label{DEF:ferchetmean}
Let $\PP$ be a probability measure on $\m$. The {population Fr\'echet function} is defined by $F(p) = \int_{\mathcal{M}} \rho^2(p, x) \, \PP(dx)$, the {population Fr\'echet mean set} of $\PP$ is the set of minimizers of $F$ and the corresponding minimum value is {population Fr\'echet variance}. Similarly, given an i.i.d. sample $X_1, \dots, X_n$, the {sample Fr\'echet mean set} is the set of minimizers of the {sample Fr\'echet function} $\hat{F}_n(p) = 1/n \sum_{i=1}^n \rho^2(p, X_i)$, and the corresponding minimum value is {sample Fr\'echet variance}.
\end{definition}


Unlike its Euclidean counterpart, the existence and uniqueness of the Fr\'echet mean require careful consideration on a general manifold. For simplicity, we assume that the support of $\PP$ lies within a convex geodesic ball, which is sufficient to ensure both existence and uniqueness. This condition is stated formally in Assumption~\ref{BPassump1}.

\begin{assumption}\label{BPassump1}
The support of the probability measure $\PP$ is contained in a closed geodesic ball $B(m_0, r)$ with some center $m_0$ and radius $r < \pi/(4\sqrt{\kappa})$, where $\kappa$ is an upper bound on the sectional curvatures of $\m$. If $\kappa \le 0$, then any $r < \infty$ suffices. 
\end{assumption}

We refer the reader to \citet{Bhattacharya2003, bhattacharya2012nonparametric} for a more comprehensive review on this topic. Under Assumption \ref{BPassump1}, the population and sample Fr\'echet means exist and are unique; we denote them by $\eta$ and $\hat\eta_n$, and  the population and sample Fr\'echet variance as $V$ and $\hat{V}_n$, respectively.

\subsection{Differential Privacy}\label{sec_dp}

\begin{definition}
    A mechanism $\mathbb{M}$ is said to satisfy $(\epsilon, \delta)$-differential privacy if, for all measurable subsets $S$ of its output space and for all neighbouring datasets $D$ and $D'$ differing in one individual, $\PP[\mathbb{M}(D) \in S] \le e^{\epsilon}\PP[\mathbb{M}(D') \in S] + \delta.$
\end{definition}

The parameters $\epsilon$ and $\delta$ jointly characterize the privacy loss, with smaller values indicating stronger protection. Since the $(\epsilon, \delta)$-DP definition is probabilistic, it is well defined on any probability space, including Riemannian manifolds equipped with the Borel $\sigma$-algebra. However, $(\epsilon, \delta)$-DP can be difficult to interpret or compose in high-dimensional settings. To address this issue, {Gaussian Differential Privacy} (GDP; \citealp{dong2022gaussian}) characterizes privacy through the trade-off between type~I and type~II errors in a binary hypothesis test distinguishing neighbouring datasets, quantified by a single privacy parameter $\mu$.

\begin{definition}
    A mechanism $\mathbb{M}$ is said to satisfy $\mu$-GDP if, for all neighbouring datasets $D, D'$, the corresponding output distributions $\mathbb{M}(D)$ and $\mathbb{M}(D')$ exhibit a Gaussian trade-off curve with privacy parameter $\mu$.
\end{definition}

Building on this framework, \citet{jiang2023gaussian} extend Gaussian differential privacy (GDP) to Riemannian manifolds, formulating their approach around a key feature of GDP: its analytic equivalence to $(\epsilon,\delta)$-differential privacy. For any $\mu>0$, a $\mu$-GDP mechanism satisfies $(\epsilon,\delta_{\mu}(\epsilon))$-DP for all $\epsilon\ge0$, for $
\delta_{\mu}(\epsilon):=\Phi(-\epsilon/\mu+\mu/2)-e^{\epsilon}\Phi(-\epsilon/\mu-\mu/2)
$, where $\Phi$ denotes the cumulative distribution function of the standard normal distribution. This leads to the definition of GDP on Riemannian manifolds as follows.

\begin{definition}\label{def_gdp_riemann}
    A $\m$-valued data-releasing mechanism $\mathbb{M}$ is said to be $\mu$-GDP if it satisfies $(\epsilon, \delta_\mu(\epsilon))$-DP for all $\epsilon \ge 0$.
\end{definition}

This formulation exploits the analytic equivalence between GDP and $(\epsilon,\delta)$-DP to give a concrete definition that avoids the technical complications of hypothesis-based formulations on manifolds. In particular, the mapping $\delta_{\mu}(\epsilon)$ provides a tractable way to express the privacy parameter $\mu$, which quantifies the Gaussian trade-off between neighboring datasets as in the Euclidean case.

In the manifold setting, the privacy guarantee depends on the sensitivity of the released statistic with respect to the Riemannian metric. Formally, for a data-dependent mapping or a {query}, $f: \mathcal{D} \to \mathcal{M}$, its global {sensitivity} is defined as
$
\Delta := \sup_{D\simeq D'} \rho\big(f(D), f(D')\big).
$

The quantity $\Delta$ measures the maximal perturbation of the output induced by replacing a single data point, capturing both the geometry of the manifold and the stability of the statistic. As in the Euclidean setting, the GDP privacy parameter $\mu$ is determined jointly by this sensitivity and the dispersion of the noise distribution. Consequently, achieving high utility requires careful calibration of the privacy budget $\mu$ against the noise dispersion parameter when enforcing differential privacy on manifolds.


\section{Theoretical Results}\label{sec3}

In this section, we extend the GDP mechanism to the setting of Riemannian manifolds and establish the corresponding theoretical framework for statistical inference based on privatized Fr\'echet mean and variance. Specifically, we consider two categories of differential privacy mechanisms, respectively defined on homogeneous manifolds with positive curvature and on Hadamard manifolds.



\subsection{DP Estimation of Fr\'echet Mean and Variance}
\label{sec_dp_estimate}



For data $D = \{X_1, \dots, X_n\}$ drawn i.i.d. from a probability measure $\mathbb{P}$ satisfying Assumption~\ref{BPassump1}, we aim to construct differentially private Fr\'echet mean estimators $\hat{\eta}_n^{\ddp}$ by injecting $\mathcal{M}$-valued random noise into the sample Fr\'echet mean $\hat{\eta}_n$.

\par The manifold-valued noise ensures that the perturbed estimator remains on the manifold, unlike methods that embed $\hat{\eta}_n$ into an ambient Euclidean space and inject Euclidean noise. Achieving differential privacy requires a careful design of the noise injection, which motivates the construction of suitable privacy mechanisms.

\begin{remark}\label{remark_sensi}
    Note that under Assumption \ref{BPassump1}, 
    the sensitivity of the sample Fr\'echet mean is, for $\kappa$ an upper bound on the sectional curvatures of $\m$,  
    \begin{equation}\label{eq_sensi}
    \Delta_\eta = 2\lambda(r, \kappa)r/n, \;\text{where} \; 
    \lambda(r, \kappa) = \begin{cases}
        r^{-1} \kappa^{-1/2}\tan(\sqrt{\kappa}2r) - 1, & \kappa > 0; \\
        1, & \kappa \le 0. \\
    \end{cases}
    \end{equation}
    See \citet{reimherr2021} for detailed derivations.
\end{remark}

We introduce two privacy mechanisms, each tailored to a particular class of manifolds. The first, referred to as the Riemannian Gaussian (RG) mechanism, achieves GDP by injecting Riemannian Gaussian noise $\mathcal{N}^{\mathrm{RG}}(\eta, \sigma)$, whose probability density function $p_{\eta,\sigma}(y)$ with respect to the Riemannian volume measure $\nu$ is 
\[
p_{\eta,\sigma}(y):= \frac{1}{Z(\eta, \sigma)} \exp\!\left\{-\frac{1}{2\sigma^2}\rho^2(y, \eta)\right\},
\]
where $\eta$ denotes the location parameter (the center), $\sigma$ is the scale parameter governing the strength of privacy protection, and $Z(\eta, \sigma)$ is the normalizing constant.

In contrast to the RG mechanism, which operates directly on the manifold, the second mechanism, referred to as the Exponential-Wrapped Gaussian (EWG) mechanism, leverages the tangent-space structure. It achieves privacy by perturbing $\hat{\eta}_n$ with EWG noise $\mathcal{N}^{\mathrm{EWG}}(p_0, \eta, \sigma)$, obtained by pushing forward the tangent-space Gaussian distribution $\mathcal{N}(\log_{p_0}\eta, \sigma^2\bm{I}_d)$ on $T_{p_0}\mathcal{M}$ through the exponential map $\exp_{p_0}$.\footnote{For the EWG distribution, we refer to the parameter $p_0, \eta, \sigma$ as the footpoint, the center, and the scale parameter, respectively.} Since this distribution is defined as the push-forward of a Gaussian law under $\exp_{p_0}$, it requires $\exp_{p_0}$ to be injective on $T_{p_0}\mathcal{M}$. This condition naturally holds on Hadamard manifolds, where each tangent space is diffeomorphic to the manifold, ensuring the global injectivity of the exponential map.

\par In our implementation, for both the RG and EWG mechanisms, we set the center to $\hat{\eta}_n$. For the EWG mechanism we set the footpoint $p_0 = m_0$, where $m_0$ is defined in Assumption~\ref{BPassump1}. For both mechanisms, the scale parameter $\sigma$ controls the trade-off between privacy and accuracy. 

\par Therefore, for these two representative classes of manifolds, we obtain the DP Fr\'{e}chet mean estimator using the corresponding privacy mechanisms, as summarized in Algorithm~\ref{alg_dp_mean}.

\begin{algorithm}
   \textbf{Input:} Sample Fr\'echet mean $\hat{\eta}_n$, data radius $r$, privacy budget $\mu$, footpoint $m_0$; \\
  \textbf{Output:} DP Fr\'echet mean estimate $\hat{\eta}_n^{\ddp}$;\\ [-1em]
  \begin{algorithmic}[1] 
    \State \textbf{Let} $\sigma_{n,\eta} = \Delta_\eta / \mu$ with $\Delta_\eta$ given in \eqref{eq_sensi}.
    \If{$\m$ is a homogeneous manifold of positive curvature}
        \State \textbf{Sample} $\hat{\eta}_n^{\ddp} \sim \N^{\RG}(\hat{\eta}_n, \sigma_{n,\eta})$. \Comment{RG mechanism}
    \ElsIf{$\m$ is a Hadamard manifold}
        \State \textbf{Sample} $\hat{\eta}_n^{\ddp} \sim \N^{\EWG}(m_0, \hat{\eta}_n, \sigma_{n,\eta})$. \Comment{EWG mechanism}
    \EndIf
    \State \textbf{Return:} $\hat{\eta}_n^{\ddp}$.
  \end{algorithmic}
    \caption{DP Fr\'echet Mean Estimation}
    \label{alg_dp_mean}
\end{algorithm}
\begin{remark}
    The footpoint $p_0$ used in the EWG mechanism can be any point within $B(m_0, r)$. However, the utility of the DP Fr\'echet mean estimator will be impacted by the location of $p_0$ and the curvature of the manifold. Ideally, one would use existing prior knowledge (if it exists) to select a footpoint near the sample Fr\'echet mean. 
\end{remark}

A central challenge in designing privacy mechanisms lies in determining the quantitative relationship between $\sigma$ and $\mu$. In the original RG mechanism of \citet{jiang2023gaussian}, this relationship is defined implicitly through an infinite collection of inequalities, which makes direct implementation difficult. To address this issue, \citet{jiang2023gaussian} employed an MCMC-based calibration procedure for $\mu$; however, this approach applies only to manifolds of constant curvature—such as spheres, Euclidean spaces, and hyperbolic spaces—and incurs substantial computational cost.

We overcome both limitations by introducing an analytical calibration method for the privacy budget. Drawing on geometric comparison theory, we derive an explicit calibration formula for homogeneous manifolds of positive curvature. Likewise, for Hadamard manifolds, the proposed EWG mechanism admits a fully analytical calibration procedure. Fixing the privacy budget $\mu$, we set $\sigma_{n,\eta} = \Delta_\eta / \mu$ for both mechanisms, rendering the overall implementation far more practical and computationally efficient. In a later section, we provide empirical comparisons showing that our derived lower bound for $\sigma_{n,\eta}$ is extremely tight. The privacy guarantee for Algorithm~\ref{alg_dp_mean} is established by the following theorem.

\begin{theorem}[Privacy Guarantee]\label{thm_dp_mean_privacy}
    Under Assumption \ref{BPassump1}, the DP Fr\'echet mean estimator $\hat{\eta}_n^{\ddp}$, under both the RG and EWG mechanisms in Algorithm \ref{alg_dp_mean}, is $\mu$-GDP.
\end{theorem}

Next, we follow the same approach by injecting noise into the sample Fr\'{e}chet variance $\hat{V}_n$. Since $\hat{V}_n \in \mathbb{R}$, Euclidean-valued noise is sufficient, and thus a traditional differential privacy (DP) mechanism can be applied. In particular, we employ the Gaussian mechanism to achieve GDP; that is, we sample
$$
\hat{V}_n^{\mathrm{DP}} \sim \mathcal{N}(\hat{F}_n(\hat{\eta}^{\ddp}_n), \sigma_{n,V}^2), \quad \hat{F}_n(p) = \frac1n \sum_{i=1}^n \rho^2 (p, X_i),  
$$
where $\sigma_{n,V}= \Delta_{V_n}/\mu$ and $
\Delta_{V_n} := \sup_{D \simeq D'} |\hat{F}_n(\hat{\eta}^{\ddp}_n) - \hat{F}_n'(\hat{\eta}^{\ddp}_n)|
$.\footnote{Here $\hat{F}_n'(\cdot)$ denotes the sample Fr\'echet function w.r.t the neighboring dataset $D'$.} We then obtain the following result for the privacy guarantee of our DP Fr\'{e}chet variance estimator. The overall procedure is summarized in Algorithm~\ref{alg_dp_var}, and its corresponding privacy guarantee is established in Theorem~\ref{thm_dp_var_privacy}.

\begin{algorithm}[h!]
   \textbf{Input:} Data $X_1, \dots, X_n$, data radius $r$, DP Fr\'echet mean $\hat{\eta}^{\ddp}_n$, privacy budget $\mu$; \\
  \textbf{Output:} DP Fr\'echet variance estimate $\hat{V}_n^{\ddp}$;\\ [-1em]
  \begin{algorithmic}[1] 
    \State \textbf{Compute} $\Delta_{V_n} = 4r^2/n$ . 
    \State \textbf{Compute} $\hat{F}_n(\hat{\eta}_n^{\ddp})= 1/n \sum_{i=1}^n \rho^2(\hat{\eta}_n^{\ddp}, X_i)$. 
    \State \textbf{Sample} $\hat{V}_n^{\ddp} \sim \N(\hat{F}_n(\hat{\eta}_n^{\ddp}), \sigma_{n,V}^2)$ with $\sigma_{n,V} = \Delta_{V_n} / \mu$. 
    \State \textbf{Return:} $\hat{V}_n^{\ddp}$.
  \end{algorithmic}
    \caption{DP Fr\'echet Variance Estimation}
    \label{alg_dp_var}
\end{algorithm}

\begin{theorem}[Privacy Guarantee for DP Fr\'echet Variance]\label{thm_dp_var_privacy}
    Under Assumption \ref{BPassump1}, given the DP Fr\'echet mean estimate $\hat{\eta}_n^{\ddp}$, the DP Fr\'echet variance estimator $\hat{V}_n^{\ddp}$ in Algorithm \ref{alg_dp_var} is $\mu$-GDP with $\Delta_{V_n} = 4r^2/n$.
\end{theorem}

\subsection{Consistency and asymptotic normality}
\label{sec_dp_clt}


 We first establish the consistency of the DP Fr\'echet mean and variance estimators.


\begin{theorem}[Consistency for DP Fr\'echet Mean]\label{thm:Consistency DP mean}
    Under Assumption \ref{BPassump1}, the DP Fr\'echet mean estimator $\hat{\eta}_n^{\ddp}$ in Algorithm \ref{alg_dp_mean} is a consistent estimator of $\eta$. Furthermore, we have $\rho(\hat{\eta}_n^\ddp, \eta) = \Oop(n^{-1/2})$.
\end{theorem}

This result is grounded in the observation that the global sensitivity of the sample Fr\'echet mean decays at the rate $\mathcal{O}(n^{-1})$, implying that the scale parameter of the differential privacy noise diminishes accordingly as the sample size $n$ increases. This asymptotic vanishing of sensitivity forces $\hat{\eta}_n^{\ddp}$ to lie in a local coordinate neighborhood of $\eta$. In particular, the intrinsic geometry of the manifold can be locally approximated by the geometry of the tangent space at $\eta$.

Next, we derive the central limit results for the DP Fr\'echet mean estimator $\hat{\eta}_n^{\ddp}$. We begin by defining the distance function with respect to the local coordinate chart $(U, \phi)$ centered at $\eta$ to be $\rho_{\phi}(u,v):=\rho(\phi^{-1}(u), \phi^{-1}(v))$, for $u,v \in \phi(U)$. Denote
\[
\Psi(u ; \theta)=\operatorname{grad}_\theta\left(\rho_\phi\right)^2(u, \theta)=\left(\frac{\partial}{\partial \theta^r}\left(\rho_\phi\right)^2(u, \theta)\right)_{r=1}^d=\left(\Psi^r(u ; \theta)\right)_{r=1}^d,
\]
we define
\vskip -10mm
\begin{equation}\label{CLTmatrix}
\bm{C} := \mathrm{Cov}\!\left[\Psi(\phi(X); \phi(\eta))\right], 
\quad
\bm{\Lambda} := \mathbb{E}\left[\partial_r \Psi^{r'}\left(\phi(X), \phi(\eta) \right)\right]_{r, r'}^d. 
\end{equation}
\begin{assumption}\label{BPassump3}
$\bm{\Lambda}$ is nonsingular and the following integrability conditions are satisfied:
\begin{align*}
&\mathbb{E}\left[ \left\| \Psi(\phi(X); \phi(\eta)) \right\| \right] < \infty, ~~ \mathbb{E}\left[ \left| \partial_r \Psi^{r'}(\phi(X), \phi(\eta)) \right| \right] < \infty \quad \text{for} \; r, r' = 1, \dots, d.
\end{align*}

\end{assumption}
Now we are ready to establish the distributional results:

\begin{theorem}[Asymptotic Normality for DP Fr\'echet Mean]\label{thm_clt_dp_mean}
    Let $\m$ be a Hadamard manifold or a homogeneous manifold of positive curvature. Under Assumptions \ref{BPassump1} and \ref{BPassump3}, for the DP Fr\'echet mean $\hat{\eta}_n^{\ddp}$ in Algorithm \ref{alg_dp_mean}, one has for $n\to \infty$
    \begin{equation}
        \left(\frac{1}{n} \bm{\Lambda}^{-1} \bm{C} \bm{\Lambda}^{-1} + \sigma_{n,\eta}^2 \bm{I}_d\right)^{-1/2}(\phi(\hat{\eta}_n^{\ddp}) - \phi(\eta)) \leadsto \mathcal{N}( \mathbf{0}, \bm{I}_d)
    \end{equation}
    where $\bm{\Lambda}, \bm{C}$ are defined in \eqref{CLTmatrix}. If $\m$ is a homogeneous manifold of positive curvature, we take $\phi := \log_{\eta}(\cdot)$. If $\m$ is a Hadamard manifold, we take $\phi := \log_{m_0}(\cdot)$, where $m_0$ is defined in Assumption \ref{BPassump1}.
\end{theorem}

Compared with the CLT in Euclidean spaces, the main difference arises from the expected Hessian matrix $\bm{\Lambda}$ of the squared distance function $\rho^2$. This matrix can be viewed as a quantity that characterizes how the manifold deviates from Euclidean geometry; it reduces to a multiple of the identity matrix when the manifold is Euclidean. For more details about BP-CLT, we refer to Section S.1.8 in the supplementary material.

As an immediate consequence of Theorem \ref{thm_clt_dp_mean}, we can construct an asymptotic confidence region for population Fr\'echet mean $\eta$ based on $\hat{\eta}_n^{\ddp}$ as follows.
\begin{corollary}[Asymptotic Confidence Region for Fr\'echet Mean]\label{coro_confidence_dp_mean}
    Under the assumptions and setting of Theorem \ref{thm_clt_dp_mean}, a DP  ($1 - \alpha$)-confidence region for $\eta$ is given by
    \begin{equation}\label{eqn_ci_mean}
        \left\{v \in \m \mid (\phi(\hat{\eta}_n^{\ddp}) - \phi(v))^\top\bm{\Gamma}_n^{-1}(\phi(\hat{\eta}_n^{ \ddp}) - \phi(v)) \le \chi^2_{d, 1-\alpha} \right\}, 
    \end{equation}
    where $\bm{\Gamma}_n := \frac{1}{n} \bm{\Lambda}^{-1} \bm{C} \bm{\Lambda}^{-1} + \sigma_{n,\eta}^2\bm{I}_d$ and $\chi^2_{d, 1 -\alpha}$ denotes the $(1 - \alpha)$-th quantile of the $\chi_d^2$ distribution. If $\m$ is a homogeneous manifold of positive curvature, we take $\phi := \log_{\hat{\eta}^{\ddp}_n}(\cdot)$. If $\m$ is a Hadamard manifold, we take $\phi := \log_{m_0}(\cdot)$. 
\end{corollary}

The CLT for Fr\'echet means on Riemannian manifolds can behave subtly, as shown by \cite{smearyCLT2019}, who identified smeariness on spheres when Assumption~\ref{BPassump1} fails. In this regime, the convergence slows to below the usual \( n^{-1/2} \) rate.
Nevertheless, in our DP framework, the convergence rate remains governed by the non-DP CLT rate, and an analogous result extends to the smeary-CLT setting (see Section S.3 of the supplementary material).

\par Next, we present the asymptotic normality for  $\hat{V}_{n}^{\ddp}$ with additional regularity conditions stated in Section S.1.10 of the supplementary material.

\vskip -5mm
\begin{theorem}[Asymptotic Normality for DP Fr\'echet Variance]\label{thm_var_clt}
    Under Assumptions \ref{BPassump1} and S.1.2, for the DP Fr\'echet variance estimator in Algorithm \ref{alg_dp_var}, one has,
    \[
    \left( \frac1n \sigma_F^2 + \sigma_{n,V}^2\right)^{-1/2} \left( \hat{V}_{n}^{\ddp} - V \right) \leadsto \mathcal{N}(0, 1)
    \]
    where $\sigma_F^2:= \operatorname{var}(\rho^2(X, \eta))$. 
\end{theorem}


Analogous to the Fr\'echet mean setting, we can construct an asymptotic confidence region for the Fr\'echet variance $V$.

\begin{corollary}[Asymptotic Confidence Interval for DP Fr\'echet Variance]\label{coro_confidence_dp_var}
    Under Assumptions \ref{BPassump1} and S.1.2, a DP $(1 - \alpha)$-confidence interval  for $V$ is given by 
    \begin{equation}\label{eqn_ci_variance}
        \left( \hat{V}_{n}^{\ddp} - z_{1 - \alpha/2}\sqrt{\frac1n \sigma_F^2 + \sigma_{n,V}^2}, \hat{V}_{n}^{\ddp} + z_{1 - \alpha/2}\sqrt{\frac1n \sigma_F^2 + \sigma_{n,V}^2} \right)
    \end{equation}
    where $z_{1 - \alpha/2}$ denotes the $(1 - \alpha/2)$-th quantile of the standard normal distribution. 
\end{corollary}


\section{Implementation}\label{sec4}
In \eqref{eqn_ci_mean} and \eqref{eqn_ci_variance}, $\Gamma_n$ and $\sigma_F^2$ are unknown and thus required to be estimated privately. We will discuss the details of the private estimation in this section.

\subsection{Estimating Limiting Variance for Fr\'echet Mean}

Recall that the limiting covariance matrix for the Fr\'echet mean in Theorem \ref{thm_clt_dp_mean} is of the form
\[
\frac{1}{n}\,\bm{\Lambda}^{-1}\bm{C}\bm{\Lambda}^{-1} + \sigma^2_{n,\eta}\bm{I}_d,
\]
where the matrices $\bm{\Lambda}$ and $\bm{C}$ are defined in \eqref{CLTmatrix}. To construct DP estimators for $\bm{\Lambda}$ and $\bm{C}$, we begin with their sample versions,
\begin{align*}
\hat{\bm{\Lambda}}_n = \frac{1}{n}\sum_{i=1}^n \left[\partial_r \Psi^{r'}\left(\phi(X_i), \phi(\hat{\eta}_n) \right)\right]_{r, r'=1}^d, \quad \hat{\bm{C}}_n =  \widehat{\cov} [\Psi(\phi(X); \phi(\eta))],
\end{align*}
where $\widehat{\cov}$ denotes the sample covariance. In particular, we note that
\begin{align*}
    \Psi(\phi(X); \phi(\eta)) =  -2[D_{\phi(\eta)}\phi^{-1}]^\dagger(\log_{\eta}(X)),
\end{align*}
where $\dagger$ denotes the adjoint operator. It follows that, $
    \hat{\bm{C}}_n = 4 L_{\phi, \eta}^\dagger \; \widehat{\cov}(\log_\eta(X)) \; L_{\phi, \eta}
$, 
where $L_{\phi, \eta} := D_{\phi(\eta)}\phi^{-1}$. 
Note that since we take $\phi := \log_{\eta}(\cdot)$ in the case of the RG mechanism for homogeneous manifolds of positive curvature, we have $L_{\phi, \eta} = \bm{I}$. On the other hand, when we take $\phi:=\log_{m_0}(\cdot)$ in the case of the EWG mechanism for Hadamard manifolds, we have $L_{\phi, \eta} = D_{\log_{m_0}(\eta)} \exp_{m_0}$. In both cases, we replace the $\eta$ in $\hat{\bm \Lambda}_n$ and $\hat{\bm C}_n$ with the DP estimator $\hat{\eta}_n^{\ddp}$:
\begin{align}
    \tilde{\bm{\Lambda}}_n = \frac{1}{n}\sum_{i=1}^n \left[\partial_r \Psi^{r'}\left(\phi(X_i), \phi(\hat{\eta}^{\ddp}_n) \right)\right]_{r, r'}^d, \quad \tilde{\bm{C}}_n = 4 L_{\phi, \hat{\eta}_n^{\ddp}}^\dagger \; \widehat{\cov}(\log_{\hat{\eta}_n^{\ddp}}(X)) \; L_{\phi, \hat{\eta}_n^{\ddp}}.
\end{align} 

Because individual observations $X_i$ enter both $\tilde{\bm{\Lambda}}_n$ and $\widehat{\cov}(\log_{\hat{\eta}_n^{\ddp}}(X))$, these statistics leak private information; consequently, additional privacy protection is required.
First, we note that $\tilde{\bm{\Lambda}}_n$ and $\widehat{\cov}(\log_{\hat{\eta}_n^{\ddp}}(X))$ are SPDMs, and the sum of two SPDMs is again an SPDM. Thus, we consider the half-vectorization of $\tilde{\bm{\Lambda}}_n$ and $\widehat{\cov}(\log_{\hat{\eta}_n^{\ddp}}(X))$ via the $\operatorname{vecd}$ operator (see \citet{Schwartzman2016lognormal} for more details).
It follows that we can obtain DP versions of the sample estimators by perturbing their half-vectorizations:
\begin{align}
\label{eqn_dp_lambda}
\hat{\bm{\Lambda}}_n^{\ddp}
&:= \operatorname{vecd}^{-1}\!\left[
\operatorname{vecd}\!\left(\tilde{\bm{\Lambda}}_n\right) +
\mathcal{N}\!\left(\bm{0},\, \sigma_{\Lambda}^2\,\bm{I}_{d(d+1)/2}\right)
\right],\\
\label{eqn_dp_C}
\hat{\bm{C}}_n^{\ddp}
&:= 4L_{\phi,\hat{\eta}_n^{\ddp}}^\dagger\operatorname{vecd}^{-1}\!\left[
\operatorname{vecd}\!\left(\widehat{\cov}(\log_{\hat{\eta}_n^{\ddp}}(X))\right) +
\mathcal{N}\!\left(\bm{0},\, \sigma_C^2\,\bm{I}_{d(d+1)/2}\right)
\right] L_{\phi,\hat{\eta}_n^{\ddp}},
\end{align}
where $\sigma_{\Lambda} = \Delta_{\Lambda}/\mu$, $\sigma_{C} = \Delta_{C}/\mu$, $\Delta_{\Lambda}$ and $\Delta_{C}$ denote the $\ell_2$ sensitivities of $\operatorname{vecd}(\tilde{\bm{\Lambda}})$, and $\operatorname{vecd}(\widehat{\cov}(\log_{\hat{\eta}_n^{\ddp}}(X)))$, respectively.

Lastly, we need to evaluate the sensitivities $\Delta_{\Lambda}$ and $\Delta_{C}$. To bound these quantities, we impose the following assumption.

\begin{assumption}\label{assump_limit_cov}
    Given $\hat{\eta}_n^{\ddp}$, $\|\log_{\hat{\eta}_n^{\ddp}}(X_i)\|_F\le R$ for all $1\le i\le n$.
\end{assumption}

Note that if Assumption \ref{BPassump1} holds, then Assumption \ref{assump_limit_cov} is satisfied with $R=2r$. Alternatively, one may set $R=r$ by truncating $\phi(X_i)$ so that $\|\log_{\hat{\eta}_n^{\ddp}}(X_i)\| \le r$.

To bound the sensitivity $\Delta_{\Lambda}$, we require the following condition on the Hessian of the squared distance function.

\begin{assumption}\label{assump_limit_hess}
Given $\hat{\eta}_n^{\ddp}$, we denote $H_i := \left[\partial_r \Psi^{r'}\left(\phi(X), \phi(\hat{\eta}_n) \right)\right]_{r, r'}^d$ and assume that $\|H_i\|_{F}\le B_H$ for $1\le i\le n$.
\end{assumption}


\begin{theorem}\label{thm_mean_limiting_covariance}
\par Under Assumptions \ref{assump_limit_cov} and \ref{assump_limit_hess}, we have $\Delta_{C} \le 6 R^2/n$ and $\Delta_{\Lambda} \le 2 B_H/n$. Furthermore, the DP limiting covariance estimator,
    \[
    \hat{\bm \Gamma}^{\ddp}_n := \frac{1}{n} \left(\hat{\bm{\Lambda}}_n^{\ddp}\right)^{-1} \hat{\bm{C}}_n^{\ddp} \left(\hat{\bm{\Lambda}}_n^{\ddp}\right)^{-1} + \sigma_{n,\eta}^2\bm{I}_d,
    \]
    together with the DP Fr\'echet mean estimate $\hat{\eta}_n^{\ddp}$ is $\sqrt{3}\mu$-GDP, where $\hat{\bm \Lambda}_n^{\ddp}$ and $\hat{\bm C}_n^{\ddp}$ are defined in \eqref{eqn_dp_lambda} and \eqref{eqn_dp_C}, respectively. Lastly, $\hat{\bm\Lambda}_n^{\ddp}$ and $\hat{\bm C}^{\ddp}_n$ are consistent estimators of ${\bm\Lambda}$ and ${\bm C}$, respectively.
\end{theorem}

To estimate the unknown limiting variance of the DP Fr\'echet variance estimator, it is sufficient to estimate $\sigma_F^2$ privately. Denote the sample estimator of $\sigma_F^2$ by
\[
\hat{\sigma}_F^2 := \frac{1}{n}\sum_{i=1}^n \rho^4\!\big(X_i,\hat{\eta}_{n}\big)
 - \left(\frac{1}{n}\sum_{i=1}^n \rho^2\!\big(X_i,\hat{\eta}_{n}\big)\right)^{\!2} = \frac{1}{n}\sum_{i=1}^n \rho^4\!\big(X_i,\hat{\eta}_{n}\big)
 - \left( \hat{V}_n\right)^{\!2} 
\]
Analogous to $\hat{V}_{n}^{\ddp}$, we apply the (Euclidean) Gaussian mechanism for GDP to obtain a DP estimator of $\sigma_F^2$. Specifically, denote 
\begin{align*}
    \tilde{\sigma}_F^2 &:= \frac{1}{n}\sum_{i=1}^n \rho^4\!\big(X_i,\hat{\eta}^{\ddp}_{n}\big)  - \left( \hat{V}^{\ddp}_n\right)^{\!2}, \\
    \Delta_\sigma &:= \sup_{D \simeq D'}\left|\frac{1}{n}\sum_{i=1}^n \rho^4\!\big(X_i,\hat{\eta}^{\ddp}_{n}\big) - \frac{1}{n}\sum_{i=1}^n \rho^4\!\big(X'_i,\hat{\eta}^{\ddp}_{n}\big)\right|,
\end{align*}
we generate the DP estimator for $\sigma^2_F$ as $\hat{\sigma}_{F}^{2,\ddp} := \tilde{\sigma}_F^2 + \mathcal{N}\!\left(0,\Delta_{\sigma}^2/\mu^2\right).
$
This yields a privatized estimator for the limiting variance, as formalized in the following result.

\begin{theorem}[Privacy and Consistency Guarantee for DP $\sigma_F^2$ Estimator]\label{Thm:dplimitvariance}
    Under Assumption \ref{BPassump1}, given the DP estimates $\hat{\eta}_n^{\ddp}, \hat{V}_n^{\ddp}$, the DP limiting variance estimator $\hat{\sigma}_F^{2,\ddp} \sim \N(\tilde{\sigma}_F^2, \sigma^2)$ with $\sigma = \Delta_{\sigma} / \mu$ is $\mu$-GDP where $\Delta_{\sigma} = 16r^4/n$. Furthermore, $\hat\sigma_F^{2, \ddp}$ is a consistent estimator of $\sigma_F^2$. 
\end{theorem}

\section{Numerical experiments}\label{sec5}
In this section, we evaluate our DP estimators and confidence regions on simulated data from the unit sphere and the SPDM space.

\par For the unit sphere $S^{d}$, a canonical example of a homogeneous manifold with positive curvature and therefore an ideal testbed for our methodology, we set $d=2$ and generate datasets $D=(X_1,\ldots,X_n)$ by sampling uniformly from the geodesic ball $B(\eta,\pi/8)\subset S^{2}$, where the centre $\eta$ is redrawn at random for each replication. For the space of symmetric positive definite matrices \(S_m^+\), with $m=2$, we generate data \(D=\{X_1,\ldots,X_n\}\) by first sampling uniformly from the geodesic ball \(B(\bm{0}, 1.5)\) on the tangent space $T_{\bm{I}_m}S_m^+$ and then map to $S_m^+$ via $\exp_{\bm{I}_m}(\cdot)$. We endow \(S_m^+\) with the Fisher--Rao affine-invariant metric, under which \(S_m^+\) forms a Hadamard manifold with strictly negative sectional curvature. We refer readers to Section~S.1.7 of the supplementary material for further details on this metric.

The sample size is fixed at $n=600$ for both the Fr\'echet mean and Fr\'echet variance, while the privacy budget varies over $\mu\in\{0.1,0.2,\ldots,2.5\}$. The number of Monte Carlo replications is $1000$.


For each privacy level, we compute the sample Fr\'echet mean \( \hat{\eta}_n \) from \(D\) using a gradient descent algorithm, together with the corresponding sample Fr\'echet variance \( \hat{V}_n \). We then construct the differentially private Fr\'echet mean estimator \( \hat{\eta}_n^{\ddp} \) and variance estimator \( \hat{V}_n^{\ddp} \) according to Algorithms~\ref{alg_dp_mean} and~\ref{alg_dp_var}. For each Monte Carlo replication, we record \( \rho(\hat{\eta}_n,\eta) \), \( \rho(\hat{\eta}_n^{\ddp},\eta) \), \( |\hat{V}_n - V| \), and \( |\hat{V}_n^{\ddp} - V| \). We summarize these errors by their Monte Carlo means, denoted MD (mean distance), and report them in Tables~\ref{sim_sphere_mean} and~\ref{sim_sphere_var}. As shown in both tables, the gap between the MDs of the DP and non-DP estimators shrinks as the privacy budget increases.

\par It is worth noting that the values $\mu^{*}$ in Table~\ref{sim_sphere_mean} are obtained using an MCMC-based procedure that computes the exact (analytical) privacy budget corresponding to a given noise level $\sigma_{n,\eta}$ and sensitivity $\Delta_{\eta}$. For a target privacy budget $\mu$, our implementation instead employs the lower bound in Algorithm~\ref{alg_dp_mean} to obtain a fast approximation to $\sigma_{n,\eta}$; we then compute the associated exact privacy budget $\mu^{*}$ as well. The results show that our lower bound is extremely close to the exact value while avoiding the substantial computational cost of the MCMC calibration, thereby making the DP inference framework more practical and efficient.

To construct confidence regions for the Fr\'echet mean, we estimate the two components of the limiting covariance, \( \hat{\boldsymbol{\Lambda}}_n \) and \( \hat{\boldsymbol{C}}_n \). The non-DP confidence region is obtained by substituting \( \boldsymbol{\Lambda} \) and \( \boldsymbol{C} \) in Corollary~\ref{coro_confidence_dp_mean} with their sample estimates \( \hat{\boldsymbol{\Lambda}}_n \) and \( \hat{\boldsymbol{C}}_n \), and by setting \( \sigma_{n,\eta}=0 \). To generate the DP confidence region, we use the differentially private estimators \( \hat{\boldsymbol{\Lambda}}_n^{\ddp} \) and \( \hat{\boldsymbol{C}}_n^{\ddp} \) defined in~\eqref{eqn_dp_lambda} and~\eqref{eqn_dp_C}, plug them into Corollary~\ref{coro_confidence_dp_mean}, and set $
\sigma_{n,\eta}=\Delta_\eta / (\mu/\sqrt{3}).
$

Similarly, for Fr\'echet variance, we split the privacy budget by allocating \( \mu/\sqrt{3} \) to \( \hat{\eta}_n^{\ddp},\hat{V}_n^{\ddp} \) and to \( \hat{\sigma}_F^{2,\ddp} \), following the corresponding theoretical guarantee. We report the finite-sample coverage of the DP confidence regions at significance level \( \alpha=0.05 \) in Tables~\ref{sim_sphere_mean}--\ref{sim_spd_var}. Empirically, the coverage rates of the DP asymptotic confidence regions for both the Fr\'echet mean and variance remain close to the nominal level across most settings. In a few cases involving the Fr\'echet variance, the coverage rate is slightly below the nominal level, which may be attributed to the fact that the estimator \( \hat{\sigma}_F^{2,\ddp} \) can occasionally underestimate \( \hat{\sigma}_F^{2} \).

\begin{figure}[h]
    \centering
    \subfigure[$\mu = 0.3$]{\includegraphics[width=0.3\textwidth]{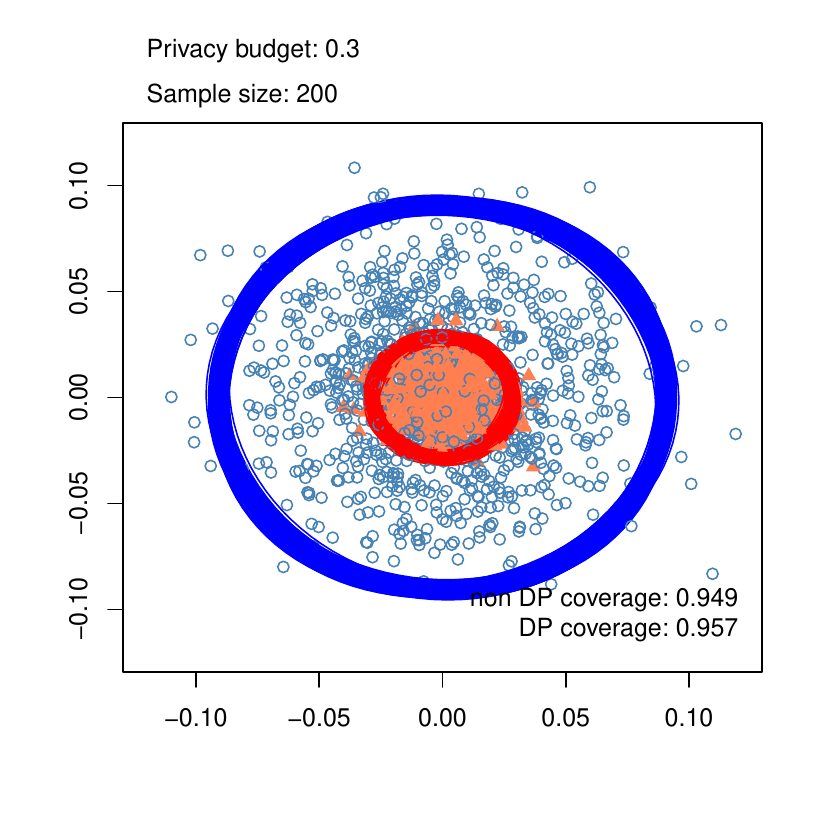}} 
    \subfigure[$\mu = 0.7$]{\includegraphics[width=0.3\textwidth]{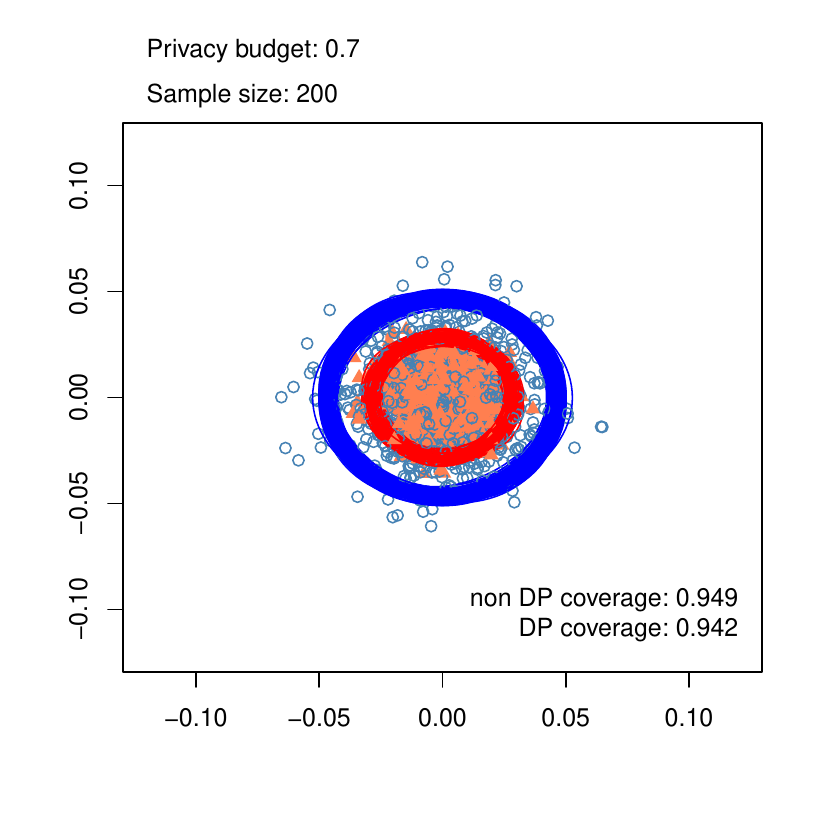}} 
    \subfigure[$\mu = 2$]{\includegraphics[width=0.3\textwidth]{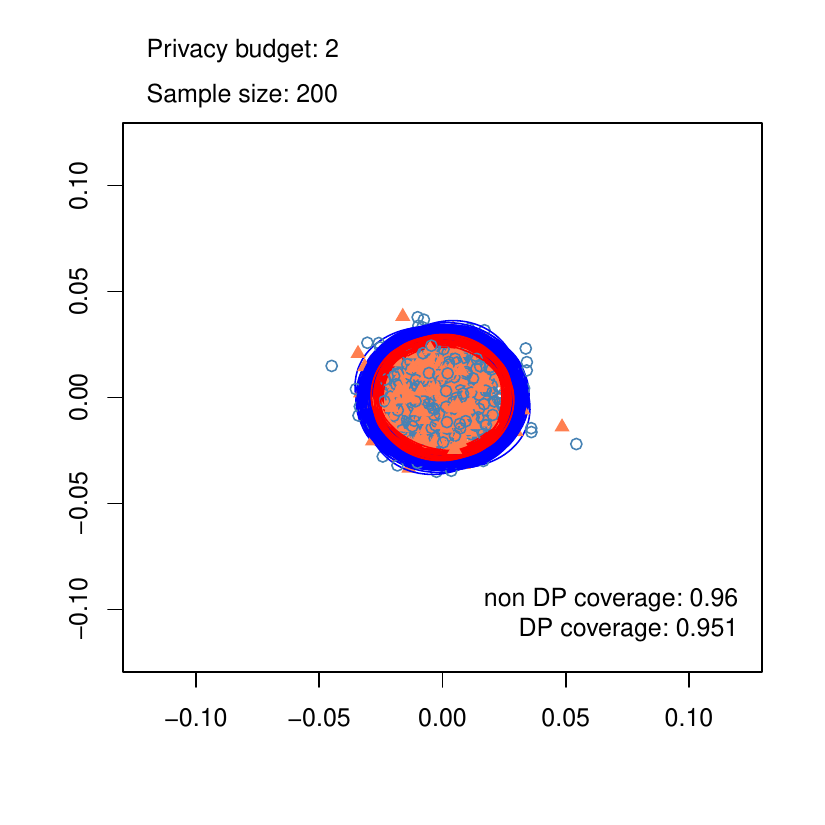}} 
    \caption{Visualization of the coverage of confidence regions. The blue ellipses and points indicate the DP confidence regions, while the red ellipses and points indicate the non-DP confidence regions.}
    \label{fig:sphere_mean_coverage}
\end{figure}

For visualization of confidence region coverage, we pull back both the DP and non-DP confidence regions to their respective tangent spaces for direct comparison. In the non-DP setting, the constructed confidence region is pulled back to the tangent space \(T_{\hat{\eta}_n}S^2\) via \(\log_{\hat{\eta}_n}\) and is represented by red ellipses in Figure~\ref{fig:sphere_mean_coverage}. The population Fr\'echet mean \(\eta\) is likewise mapped to \(T_{\hat{\eta}_n}S^2\) as \(\log_{\hat{\eta}_n}\eta\), shown as triangular red points.

In the DP setting, the DP confidence region and the population Fr\'echet mean are pulled back to \(T_{\hat{\eta}_n^{\ddp}}S^2\) via \(\log_{\hat{\eta}_n^{\ddp}}\), and are visualized as blue ellipses and circular blue points, respectively. As illustrated in Figure~\ref{fig:sphere_mean_coverage}, the DP confidence regions progressively approach their non-DP counterparts as the privacy budget \(\mu\) increases.

\begin{table}[h]
    \centering
    \caption{Fr\'echet mean results on the sphere. Top row: MD of DP sample Fr\'echet mean. Bottom row: Coverage probabilities of DP confidence regions for Fr\'echet mean.}
    \label{sim_sphere_mean}
    \begin{tabular}{lcccccccccc}
        \toprule
        \textbf{$\mu$} & 0.1 & 0.2 & 0.3 & 0.5 & 0.7 & 1 & 1.5 & 2 & 2.5 & non-DP\\ \midrule
         \textbf{$\mu^*$} & \small 0.099 & \small 0.199 & \small 0.300 & \small 0.500 & \small 0.700 & \small 0.999 & \small 1.500 & \small 2.000 & \small 2.500& --\\
         \textbf{MD}({\small $10^{-4}$}) & 183 & 126 & 116 & 106 & 106 & 107 & 105 & 105 & 103 & 103 \\
        \textbf{coverage} & 0.941 & 0.953 & 0.945 & 0.96 & 0.954 & 0.939 & 0.941 & 0.954 & 0.96 & 0.954 \\ \bottomrule
    \end{tabular}
\end{table}

\begin{table}[h]
    \centering
    \caption{Fr\'echet variance results on the sphere. Top row: MD of DP sample Fr\'echet variance. Bottom row: Coverage probabilities of DP confidence regions for Fr\'echet variance.}
    \label{sim_sphere_var}
    \begin{tabular}{llllllllllll}
        \toprule
        \textbf{$\mu$} & 0.1 & 0.2 & 0.3 & 0.5 & 0.7 & 1 & 1.5 & 2 & 2.5 & non-DP \\ \midrule 
        \textbf{MD}({\small $10^{-5}$}) & 137 & 73.9 & 51.1 & 31.8 & 25.6 & 21.5 & 17.6 & 16.2 & 16.4 & 14.6 \\
         \textbf{coverage} & 0.957 & 0.944 & 0.928 & 0.941 & 0.931 & 0.924 & 0.938 & 0.942 & 0.931 & 0.947 \\ \bottomrule
    \end{tabular}
\end{table}

\begin{table}[h]
    \centering
    \caption{Fr\'echet mean results on the SPDM space. Top row: MD of DP sample Fr\'echet mean. Bottom row: Coverage probabilities of DP confidence regions for Fr\'echet mean.}
    \label{sim_spd_mean}
    \begin{tabular}{lcccccccccc}
        \toprule
        \textbf{$\mu$} & 0.1 & 0.2 & 0.3 & 0.5 & 0.7 & 1 & 1.5 & 2 & 2.5 & non-DP\\ \midrule 
        \textbf{MD}({\small $10^{-3}$}) & 144 & 78.1 & 59.9 & 46.6 & 42.1 & 40.1 & 39.4 & 38.2 & 38 & 37.7 \\
         \textbf{coverage} & 0.945 & 0.943 & 0.942 & 0.95 & 0.96 & 0.966 & 0.966 & 0.962 & 0.964 & 0.971 \\\bottomrule
    \end{tabular}
\end{table}

\begin{table}[h]
    \centering
    \caption{Fr\'echet variance results on the SPDM space. Top row: MD of DP sample Fr\'echet variance. Bottom row: Coverage probabilities of DP confidence regions for Fr\'echet variance.}
    \label{sim_spd_var}
    \begin{tabular}{llllllllllll}
        \toprule
        \textbf{$\mu$} & 0.1 & 0.2 & 0.3 & 0.5 & 0.7 & 1 & 1.5 & 2 & 2.5 & non-DP \\ \midrule 
        \textbf{MD}({\small $10^{-4}$}) & 207 & 105 & 71.6 & 46.4 & 34.9 & 29.2 & 22.9 & 21.4 & 20.8 & 19.7 \\
         \textbf{coverage} & 0.95 & 0.964 & 0.939 & 0.948 & 0.937 & 0.937 & 0.946 & 0.947 & 0.95 & 0.954 \\ \bottomrule
    \end{tabular}
\end{table}

\section{Applications to Real Datasets}\label{sec6}
\subsection{Medical image data}

The first dataset, \textsf{OCTMNIST}, is from the \textsf{MedMNIST} collection \citep{Yang2021MedMNISTV1,Yang2021MedMNISTV2}, which contains $28\times 28$ grayscale OCT images resized from $109{,}309$ valid retinal OCT scans. After preprocessing, each image is mapped to a $5\times 5$ SPD covariance descriptor in $S_5^+$ (see Section~S.3 in the supplementary material).

The \textsf{OCTMNIST} dataset contains four classes, labeled $0$--$3$. For each class, we construct DP and non-DP $(1-\alpha)$ confidence regions for the Fr\'echet mean with $\alpha=0.05$. We summarize the estimated asymptotic covariance $\hat{\bm\Gamma}_n^{\ddp}$ by its three largest eigenvalues and the explained proportion $(\lambda_1+\lambda_2+\lambda_3)/\mathrm{tr}(\hat{\bm\Gamma}_n^{\ddp})$, and report the effective radius $(\mathrm{tr}(\hat{\bm\Gamma}_n^{\ddp}))^{1/2}$, volume $(\det(\hat{\bm\Gamma}_n^{\ddp}))^{1/2}$, and the distortion $d(\hat{\eta}_n,\hat{\eta}_n^{\ddp})$. Table~\ref{octmnist_mean} reports the results for Class~0; results for the remaining classes are given in Section~S.3 of the supplementary material.

\begin{figure}[h]
    \centering
    \includegraphics[width=0.23\textwidth, trim = 22cm 5cm 20cm 5cm, clip]{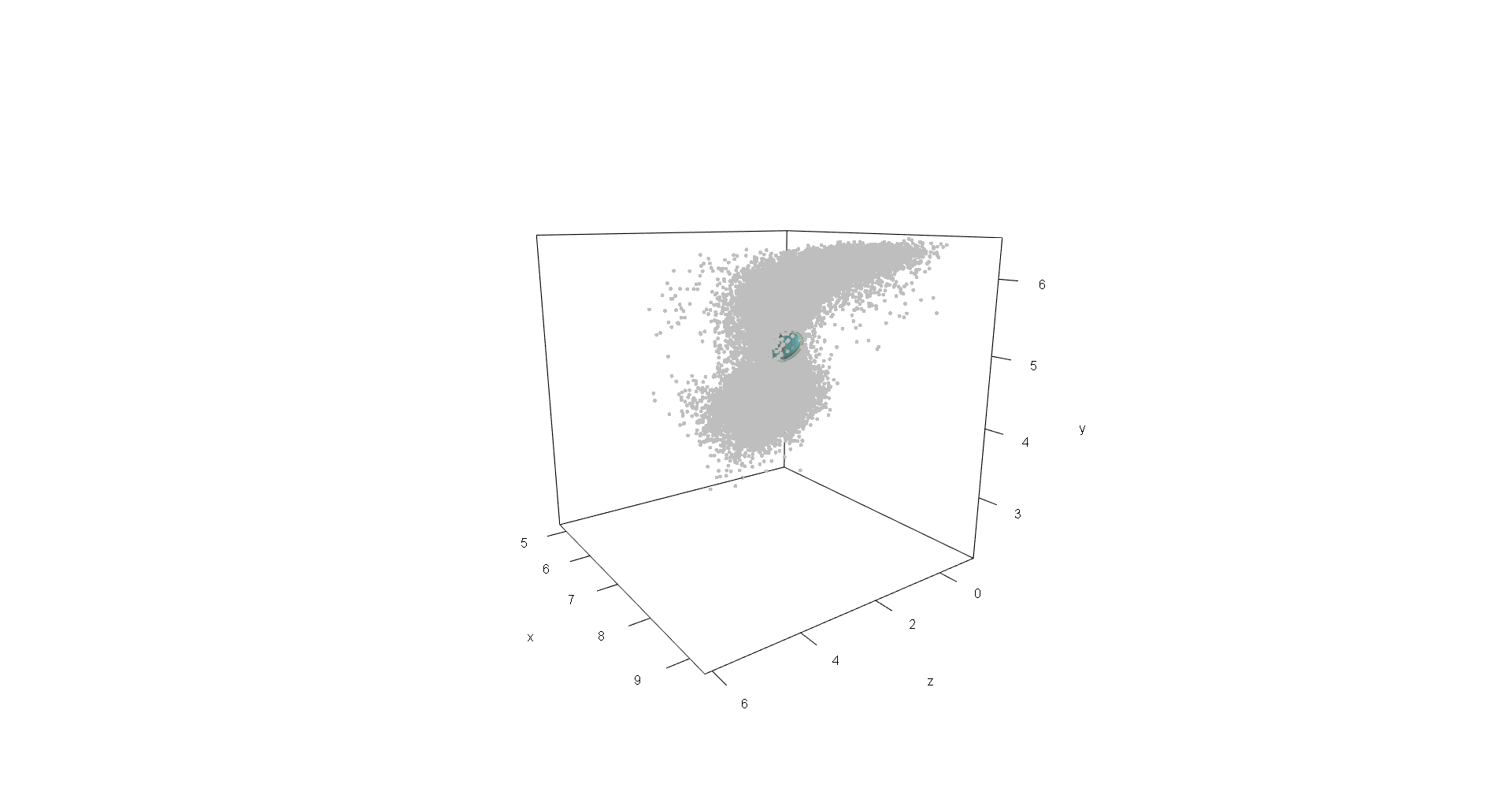}
    \includegraphics[width=0.23\textwidth, trim = 22cm 5cm 20cm 5cm, clip]{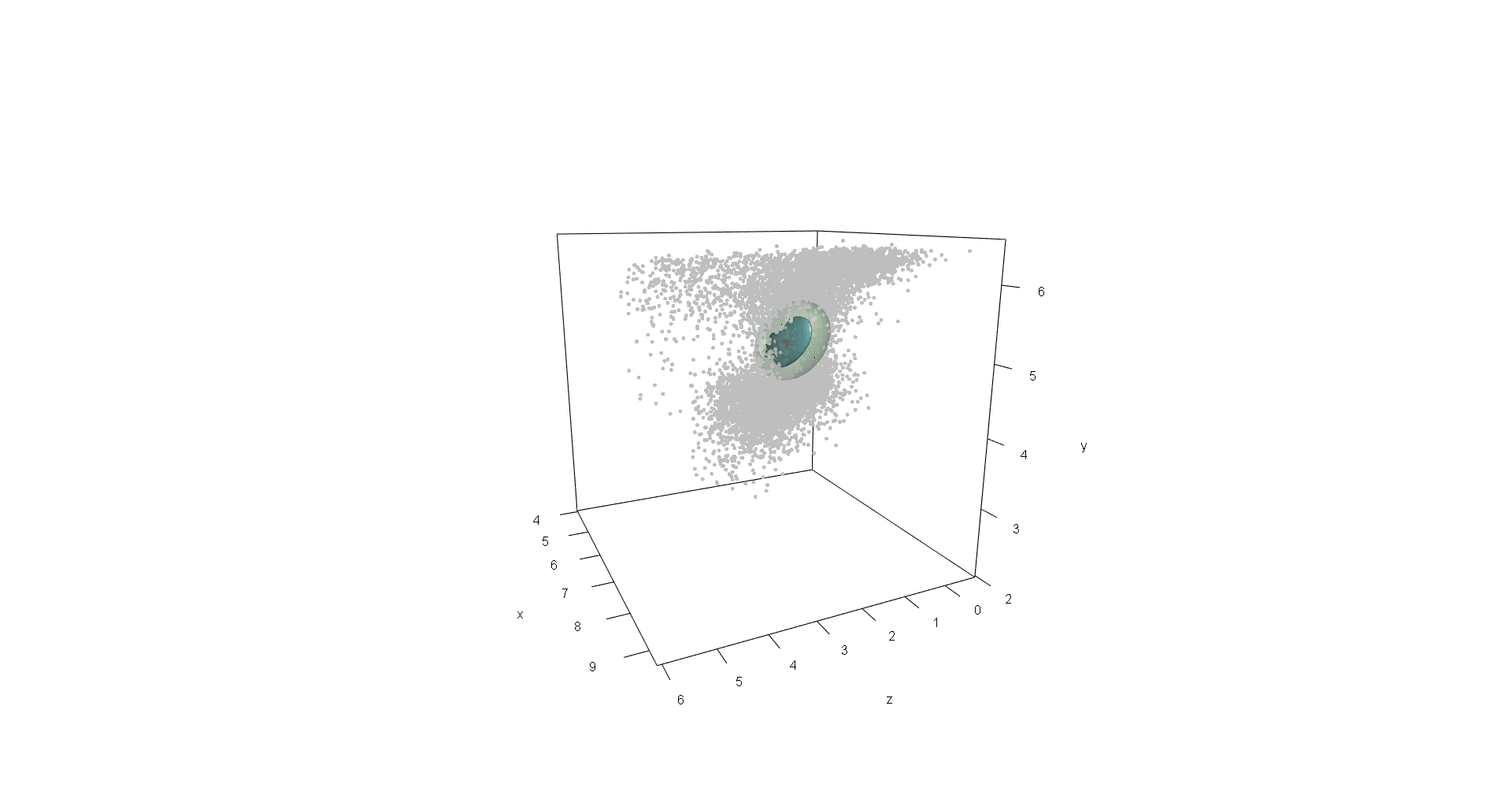}
    \includegraphics[width=0.23\textwidth, trim = 22cm 5cm 20cm 5cm, clip]{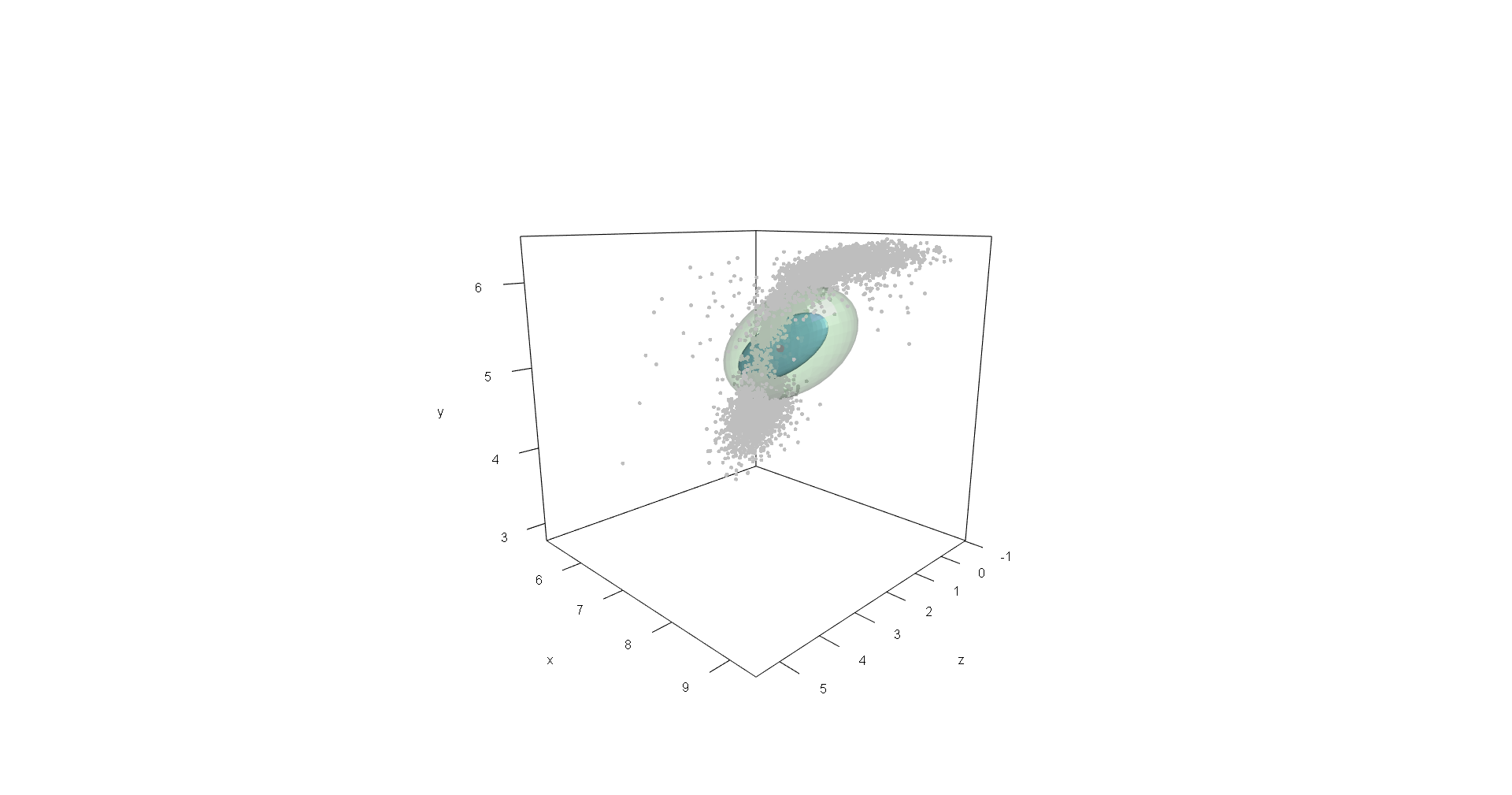}
    \includegraphics[width=0.23\textwidth, trim = 22cm 5cm 20cm 5cm, clip]{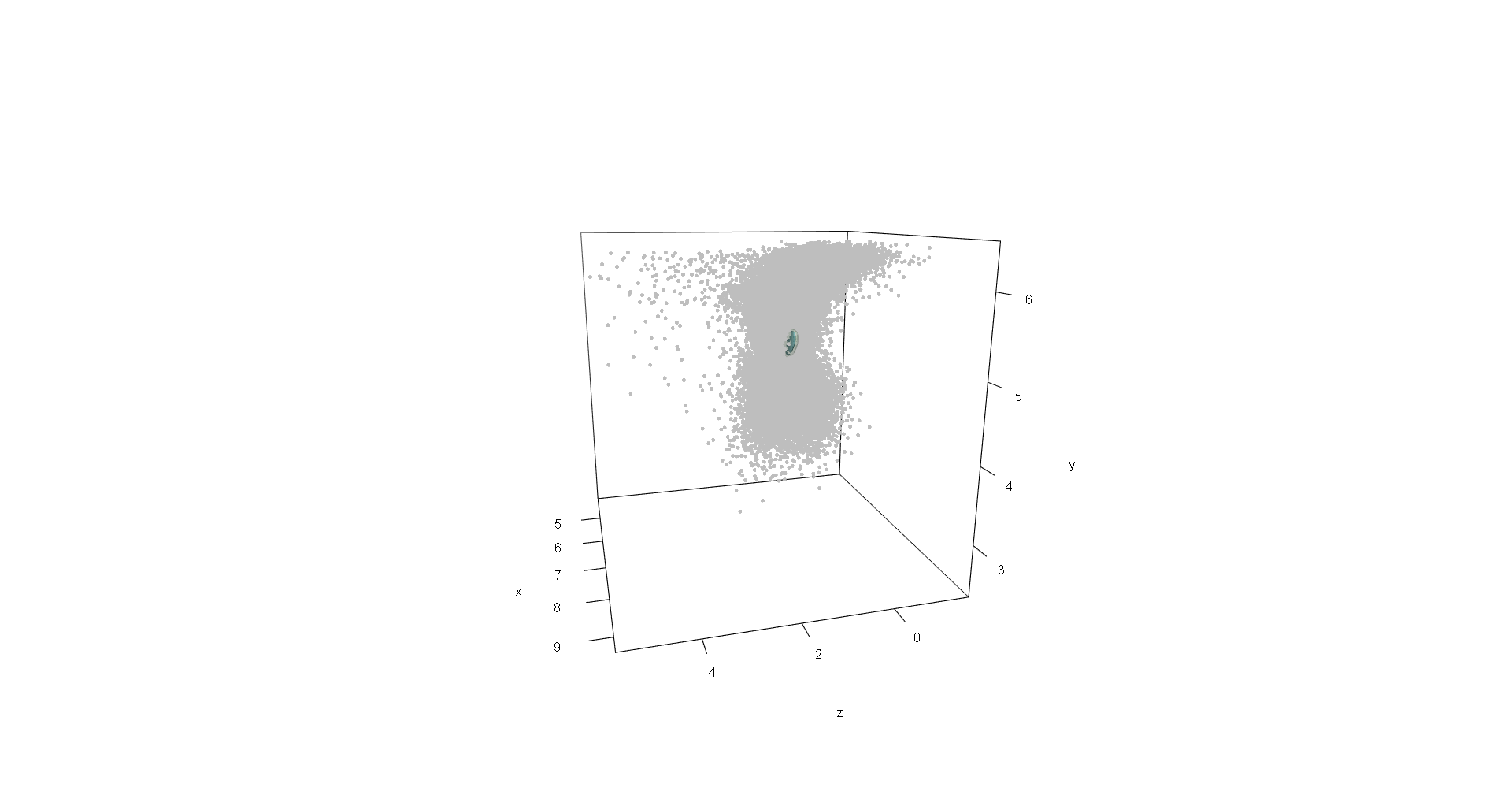}
    \subfigure[Class 0]{
    \includegraphics[width=0.23\textwidth, trim = 22cm 5cm 20cm 5cm, clip]{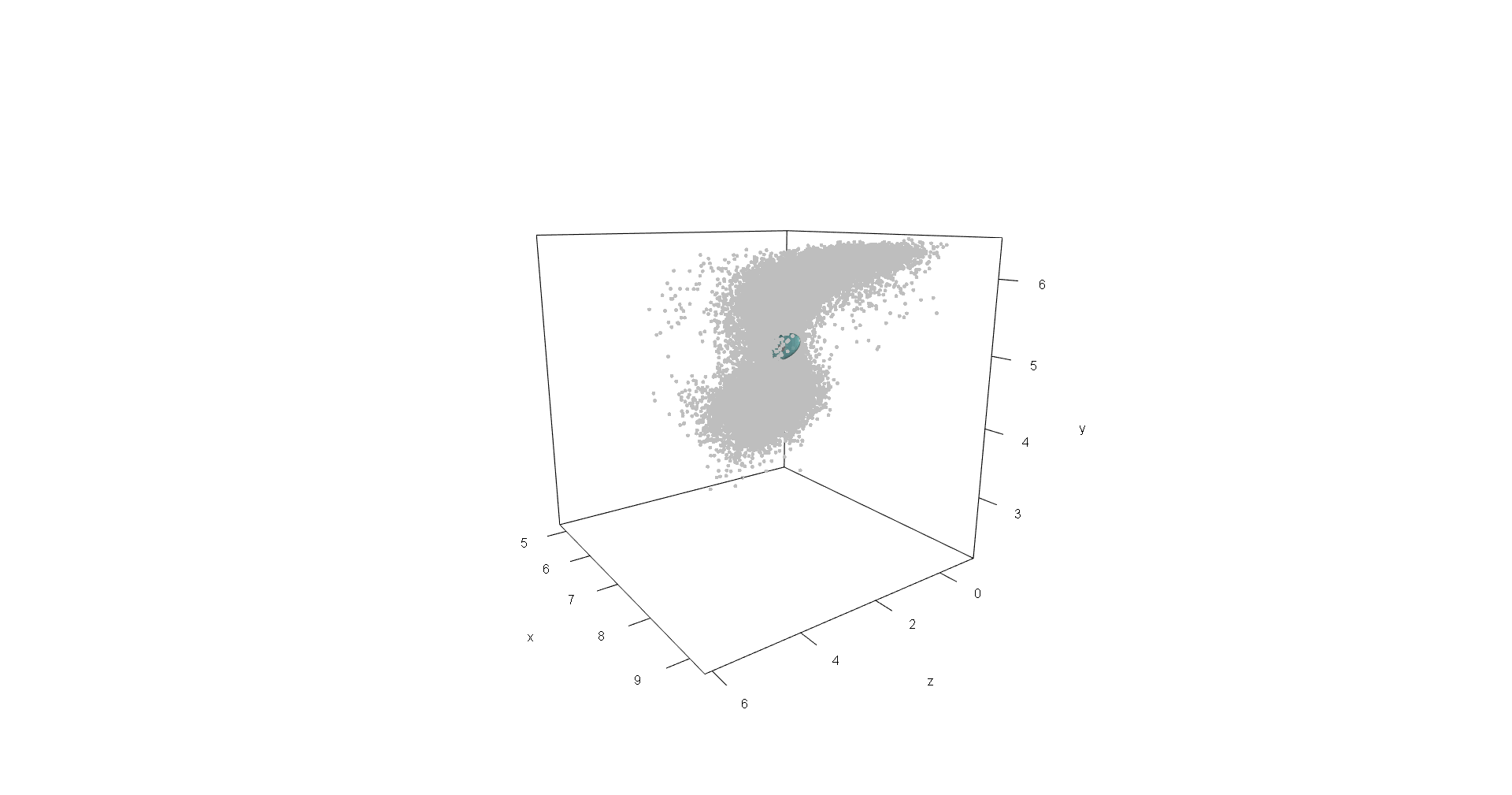}}
    \subfigure[Class 1]{
    \includegraphics[width=0.23\textwidth, trim = 22cm 5cm 20cm 5cm, clip]{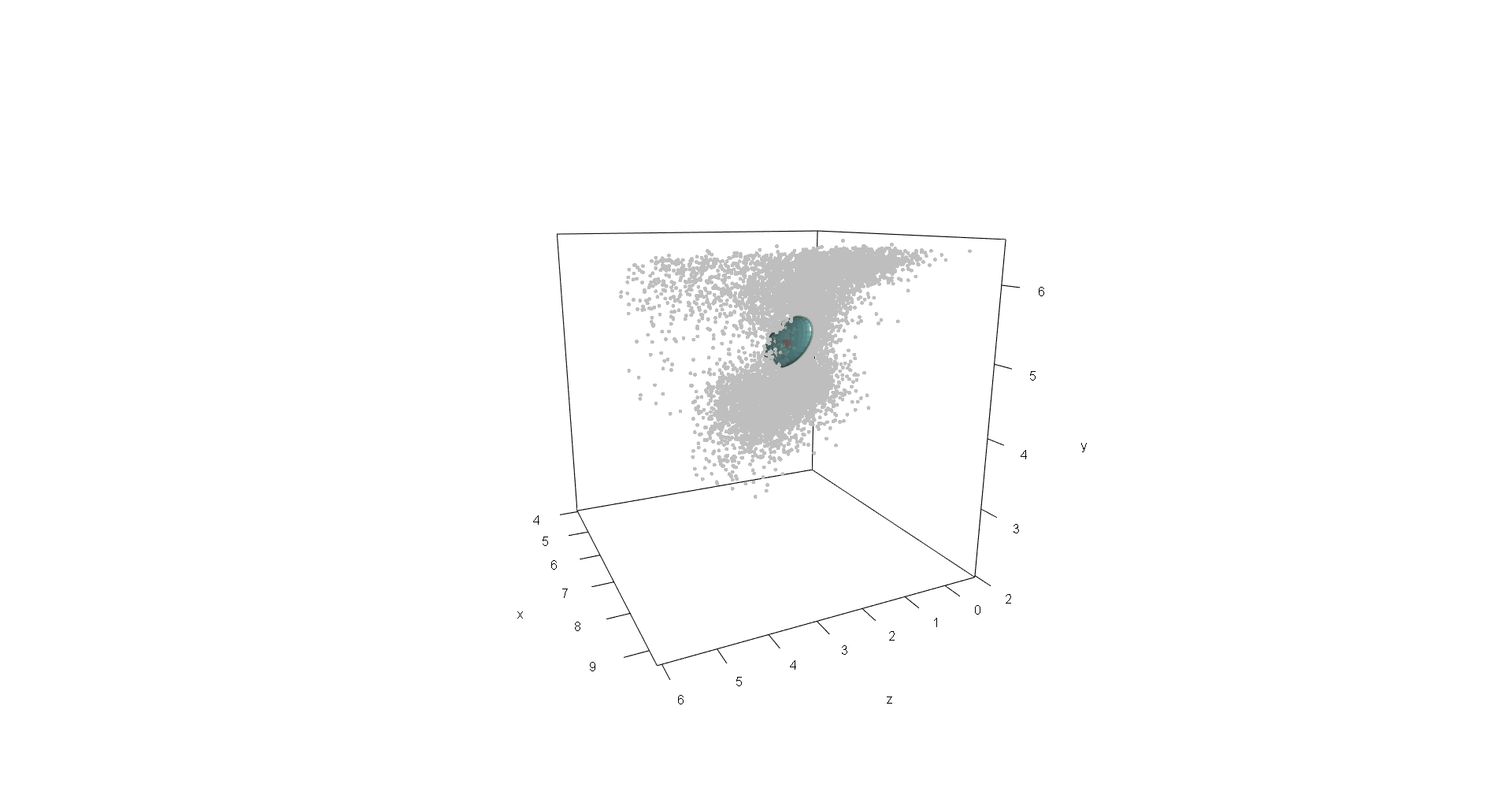}}
    \subfigure[Class 2]{
    \includegraphics[width=0.23\textwidth, trim = 22cm 5cm 20cm 5cm, clip]{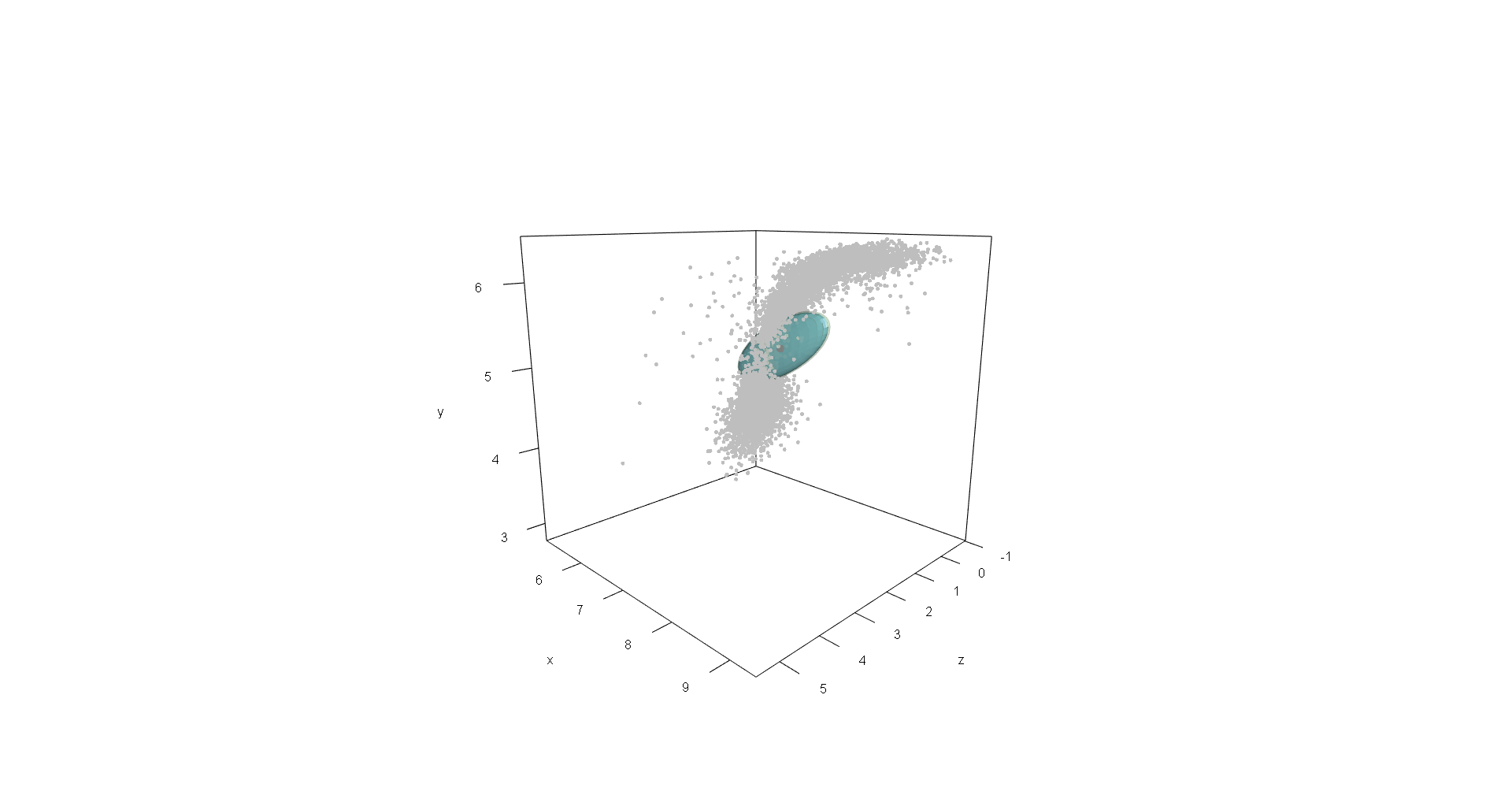}}
    \subfigure[Class 3]{
    \includegraphics[width=0.23\textwidth, trim = 22cm 5cm 20cm 5cm, clip]{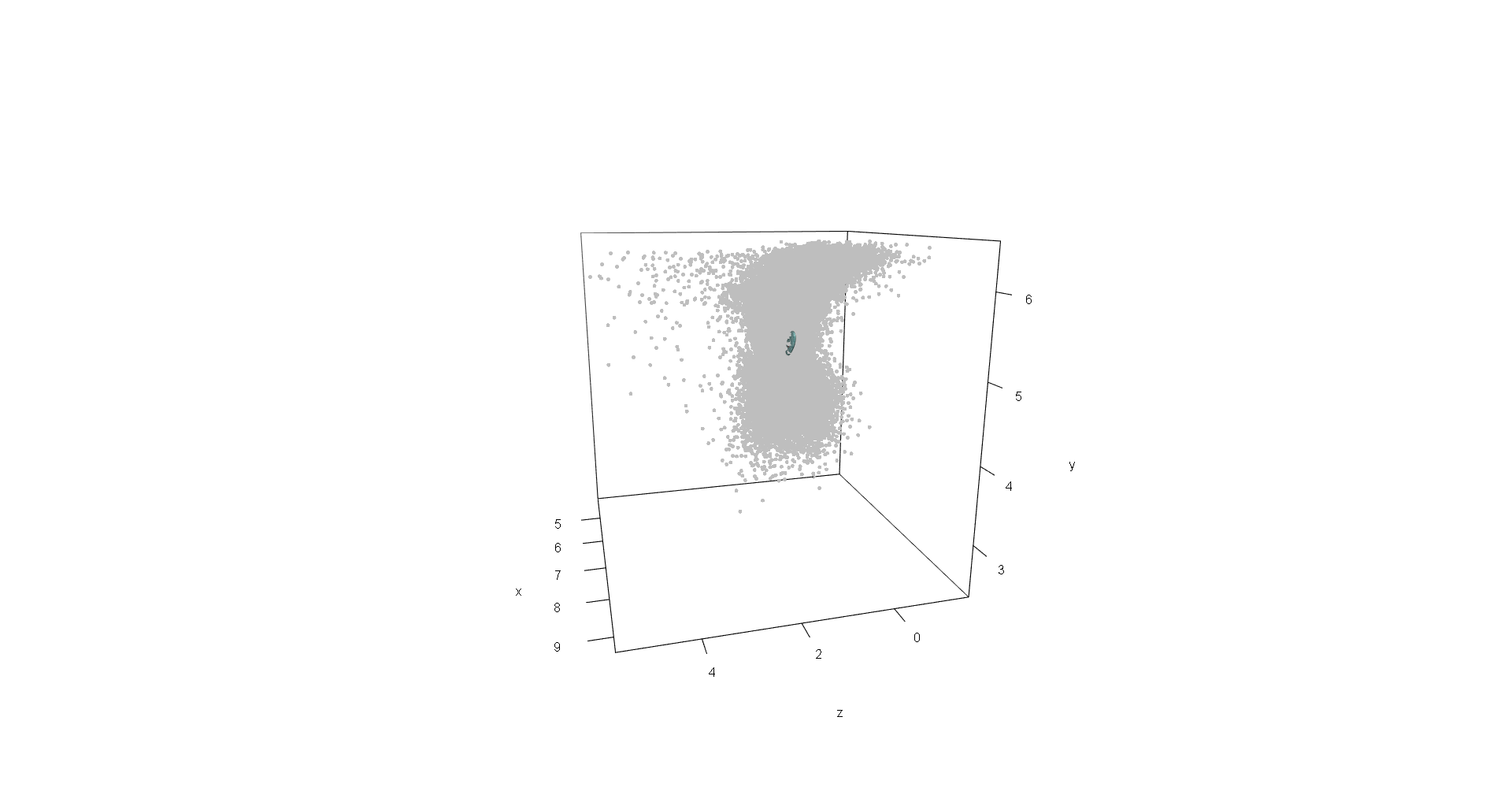}}
    \caption{DP (green) and non-DP (blue) confidence regions projected to the tangent space at $\bm I_5$ using the first three coordinates. The top and bottom rows correspond to privacy budgets $\mu=0.1$ and $\mu=1$, respectively.}
    \label{fig_octmnist_confidence_region}
\end{figure}

For visualization, we project each confidence ellipsoid onto the first three coordinates of $T_{\bm I_5}S_5^+$ and plot the resulting 3D slices (Figure~\ref{fig_octmnist_confidence_region}). Moderate privacy ($\mu=1,2$) yields DP regions close to their non-DP counterparts, whereas strong privacy ($\mu=0.1$) inflates the region and reduces the explained proportion, indicating that uncertainty spreads beyond the first three coordinates. Classes with tighter embeddings (e.g., Class~0 and Class~3) exhibit smaller regions, reflecting lower within-class variability under this SPD representation.

\begin{table}[h]
    \centering
    \caption{Results for Class $0$ in the \textsf{OCTMNIST} dataset. See Section S.3 of the supplementary material for the results of other classes.}
    \label{octmnist_mean}
    \begin{tabular}{lccccccc}
        \toprule
         & $\lambda_1$ & $\lambda_2$
 & $\lambda_3$ & ratio & rad & vol & dist \\ \midrule
        0.1 & 0.029 & 0.008 & 0.005 & 0.714 & 0.242 & 3.32 \tiny $(10^{-21})$ & 0.139\\ 
        1 &  0.026 & 0.005 & 0.002 & 0.928 & 0.190 & 2.60 \tiny $(10^{-31})$ & 0.014 \\
        2 & 0.026 & 0.005 & 0.002 & 0.935 & 0.189 & 2.63 \tiny $(10^{-33})$ & 0.007 \\
        non-DP & 0.026 & 0.005 & 0.002 & 0.942 & 0.188 & 1.96 \tiny$(10^{-39})$ & \\ \bottomrule
    \end{tabular}
\end{table}

\subsection{Sociology data}

The second dataset is the sociology data of \citet{fisher1987spherical}. In a study of 48 individuals' attitudes toward 16 occupations, judgments were made according to four criteria: Earnings, Social Status, Reward, and Social Usefulness. This produced four samples of 48 multivariate observations. Using external analysis, each observation was reduced to a unit (spherical) vector, yielding four samples of unit vectors corresponding to the four criteria.

\begin{figure}[h]
    \centering
    \includegraphics[width=0.19\textwidth, trim = 8cm 8cm 8cm 8cm, clip]{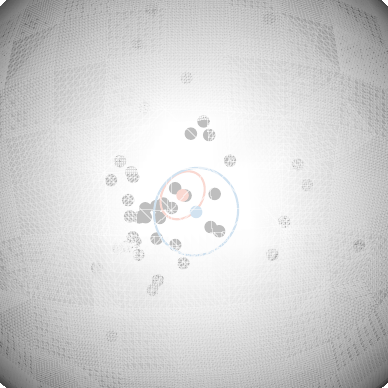}
    \includegraphics[width=0.19\textwidth, trim = 8cm 8cm 8cm 8cm, clip]{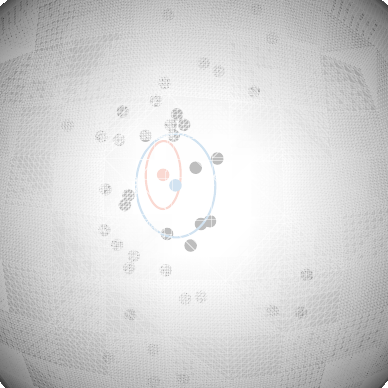}
    \includegraphics[width=0.19\textwidth, trim = 8cm 8cm 8cm 8cm, clip]{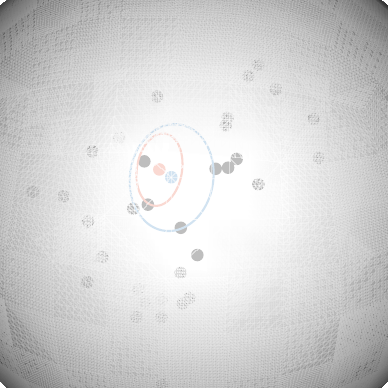}
    \includegraphics[width=0.19\textwidth, trim = 8cm 8cm 8cm 8cm, clip]{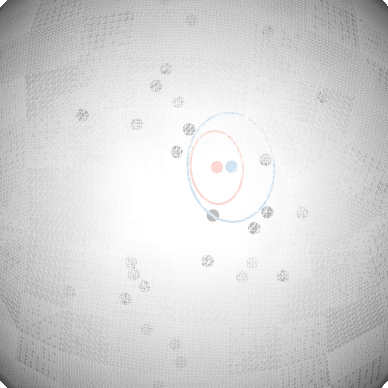}
    \includegraphics[width=0.19\textwidth, trim = 8cm 8cm 8cm 8cm, clip]{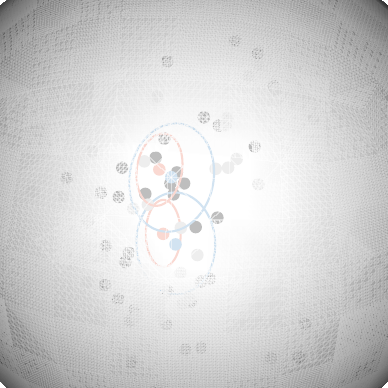}
    \subfigure[Earnings]{\includegraphics[width=0.19\textwidth, trim = 8cm 8cm 8cm 8cm, clip]{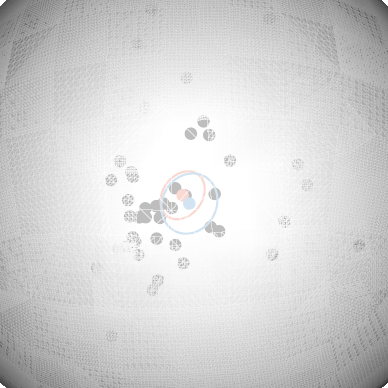}} 
    \subfigure[Social Status]{\includegraphics[width=0.19\textwidth, trim = 8cm 8cm 8cm 8cm, clip]{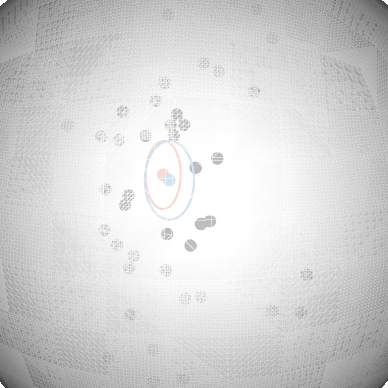}} 
    \subfigure[Reward]{\includegraphics[width=0.19\textwidth, trim = 8cm 8cm 8cm 8cm, clip]{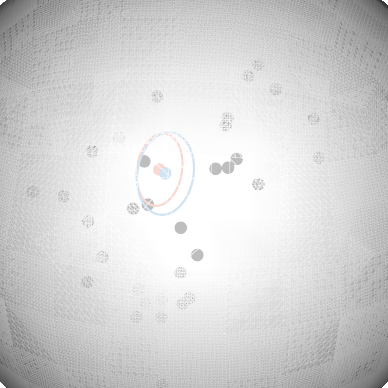}}
    \subfigure[Usefulness]{\includegraphics[width=0.19\textwidth, trim = 8cm 8cm 8cm 8cm, clip]{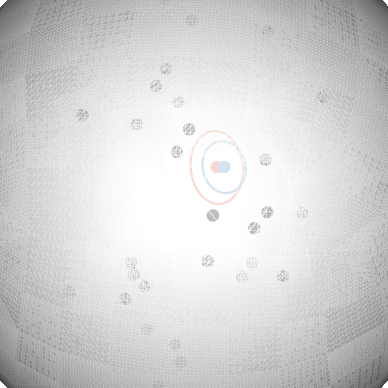}}
    \subfigure[Comparison]{\includegraphics[width=0.19\textwidth, trim = 8cm 8cm 8cm 8cm, clip]{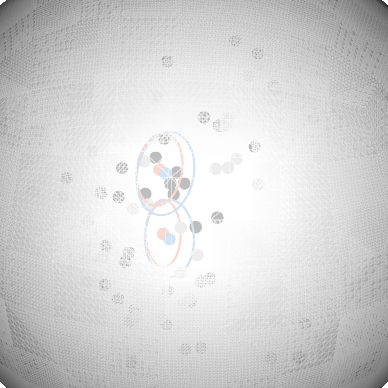}}
    \caption{DP asymptotic confidence regions constructed on sociology data. The region outlined by the red ellipse indicates the non-DP confidence region, while the blue ellipse indicates the DP confidence region. The top and bottom rows correspond to privacy budgets $\mu = 1, 2$, respectively.}
    \label{fig:occupation_confidence_region}
\end{figure}

For each criterion, we construct both non-DP and DP asymptotic confidence regions for the Fr\'echet mean. For the DP regions, we truncate the data to lie within the geodesic ball centred at the sample Fr\'echet mean with radius $\pi/8$, and display the resulting 95\% confidence regions for privacy budgets $\mu=1$ and $\mu = 2$ in Figure~\ref{fig:occupation_confidence_region}. Consistent with the behaviour observed in Section~\ref{sec5}, the DP confidence regions contract towards their non-DP counterparts as the privacy budget increases. For ease of comparison across criteria, the results for Social Status and Reward are plotted together in the last column of Figure~\ref{fig:occupation_confidence_region}. In the original non-private analysis, the 95\% regions for these two criteria overlap only slightly, indicating that the latent patterns of occupational evaluation are similar but not identical, nearby in direction, yet compatible with distinct underlying populations \citep[Chapter~7]{fisher1987spherical}. Under $\mu=2$, the DP regions are very close to the non-DP ones, and the limited overlap again suggests related but genuinely different directional patterns. When $\mu$ is reduced to $1$, the DP regions expand and overlap more, reflecting increased uncertainty under the stronger privacy constraint; nevertheless, each DP Fr\'echet mean still lies outside the other's 95\% DP region, so the central directions remain separated. In practical terms, if the two regions were to overlap almost completely, we would regard the latent occupational evaluation patterns as essentially the same; even under stronger privacy, however, they remain sufficiently distinct to support a meaningful directional difference between the two criteria.

\section{Concluding remarks}

This work addresses the gap at the intersection of geometry and privacy by developing principled, implementable differential privacy mechanisms and inference procedures for manifold-valued data, with a focus on Fr\'{e}chet mean and variance on widely used spaces such as Hadamard manifolds or homogeneous manifolds with positive curvature. By tying noise calibration to intrinsic geometric sensitivity, our framework preserves the spirit of Euclidean DP while respecting curvature and the lack of linear structure, and it delivers root-$n$ consistency, CLTs, and smeary asymptotics that enable valid uncertainty quantification under privacy constraints. The resulting estimators are fully practical, avoiding bespoke MCMC or opaque budget inversions, and thus suitable for sensitive applications in the real world.

Despite these advances, our framework has limitations. Our methods are formulated under central DP with a trusted aggregator and thus do not provide local privacy against the curator.  Extending DP inference to a local model on manifolds presents additional challenges, owing to the nonlinear geometry.  Addressing these issues represents an interesting direction for future work.




\begin{appendices}
\section{Notations}

\begin{table}[htbp!]
\centering
\caption{Notation table}
\resizebox{0.9\textwidth}{!}{
\begin{tabular}{ll}
\toprule
\textbf{Symbol} & \textbf{Description} \\
\midrule
\( \mathcal{M} \) & Riemannian manifold of dimension $d$\\
\( g \) & Riemannian metric (inner product on tangent spaces) \\
\( \rho(p, q) \) & Geodesic distance between points \( p, q \in \mathcal{M} \) \\
\( \nu \) & Riemannian volume measure on \( \mathcal{M} \) \\
\( \exp_p \)$(\log_p)$ & Riemannian exponential (respectively, logarithmic) map at point \( p \) \\
\( \Exp \)$(\Log)$ & Matrix exponential(logarithm)\\
\( B(p, r)\) & Geodesic ball centered at $p \in \m$ with radius $r$, \(\{x \in \m \mid \rho(x, p) \le r\}\) \\
\( \eta_\PP \) & Population Fr\'echet mean w.r.t. distribution \( \PP \) \\
\( \hat{\eta}_n \), \( \hat{V}_n \) and \( \hat{F}_n \)& Sample Fr\'echet mean, variance and function \\
\( \hat{\eta}_n^{\ddp} \) & DP sample Fr\'echet mean \\
\( \{X_i\}_{i=1}^n \) & Confidential data sample from some unknown distribution \( \PP \) on $\m$ \\
\( \sigma \) & Rate parameter in Gaussian or Laplace mechanism \\
\( \mu \) & Privacy budget for GDP\\
\( S_m^+ \) & Space of \( m \times m \) SPDM \\
\( \rho^{\AI} \) & Affine-Invariant distance on $S_m^+$\\
\( S_m\) & Space of \( m \times m \) symmetric matrices \\
\( \vecd \) & Vectorization map from $S_m$ to $\R^d$ \\
\( \Sigma \) and \( \hat{\Sigma} \) & Asymptotic covariance matrix and its sample estimator \\
\( D_x f \) & Differential of map $f$ at $x$ \\
\( \Hess_x f\) & Hessian of map $f$ at $x$ \\
\text{Supp}($\mathbb{P}$) & Support of the probability measure $\mathbb{P}$\\
$C(p)$ & The cut locus of $p$ \\
\bottomrule
\end{tabular}
}
\label{tab:notation}
\end{table}

\section{Supplementary material}
\par The supplementary material contains background material of Riemannian manifolds, additional results for data applications and detailed proofs of the theoretical results.

\end{appendices}


\bibliographystyle{abbrvnat}
\bibliography{bib_jrssb}
\end{document}